\documentclass[aps,12pt]{revtex4}
\usepackage{epsf}
\newcommand{\CA}{{\cal A}}

\newcommand{\CD}{{\cal D}}
\newcommand{\CE}{{\cal E}}
\newcommand{\CM}{{\cal M}}

\newcommand{\CB}{{\cal B}}
\newcommand{\CC}{{\cal C}}
\newcommand{\CF}{{\cal F}}
\newcommand{\CO}{{\cal O}}
\newcommand{\CT}{{\cal T}}
\begin{document}
\def\bp{{\mbox{\bf p}}}
\def\bP{{\mbox{\bf P}}}
\def\bD{{\mbox{\bf D}}}
\def\bk{{\mbox{\bf k}}}
\def\br{{\mbox{\bf r}}}
\def\bq{{\mbox{\bf q}}}
\def\bn{{\mbox{\bf n}}}
\def\ba{{\mbox{\bf a}}}
\def\bb{{\mbox{\bf b}}}
\def\bc{{\mbox{\bf c}}}
\def\bxi{{\mbox{\boldmath$\xi$}}}
\def\bsigma{{\mbox{\boldmath$\sigma$}}}
\def\1s0{{^1\!S_0^{++}}}
\def\sqsf{{\sqrt{s_f}}}
\thispagestyle{empty}
\begin{center}
{ \Large \bf
Exclusive charge exchange reaction $pD\to n(pp)$ within the
 Bethe-Salpeter formalism
}\\[10mm]
\noindent
L.P. Kaptari $^{a,b}$, B. K\"ampfer $^{a}$,  
S.S. Semikh $^{a,b}$, S.M. Dorkin $^{c}$\\[4mm]
\vskip 5mm
$^a$ Forschungszentrum Rossendorf, Institute for Nuclear and Hadron Physics,\\
PF 510119, 01314 Dresden, Germany\\
$^b$ Bogoliubov Laboratory of Theoretical Physics, JINR Dubna,\\
P.O. Box 79, Moscow, Russia\\
$^c$ Skobeltsyn Nuclear Physics Institute, Moscow State University, Dubna,
Russia
\end{center}
\vskip 10mm

\begin{abstract}
\noindent
The exclusive charge exchange reaction  $pD\to n(pp)$
at intermediate and high energies is studied
within the Bethe-Salpeter formalism.
The final state interaction in the detected
$pp$ pair at nearly zero excitation energy  
is described by the $^1S_0$ component of the Bethe-Salpeter
amplitude. Results of numerical calculations of polarization
observables and differential cross-section persuade that,
as in the non-relativistic case, this reaction (i) can be 
utilized as a ``relativistic deuteron polarimeter''
and (ii) delivers further  information about the elementary nucleon-nucleon
charge-exchange amplitude.
\end{abstract}

\maketitle

\section{Introduction}
\label{introd}

The investigation of polarization observables in electromagnetic and
hadronic processes at high energies provides
refinement of the information about strong interaction at
short distances and the relevant reaction mechanisms. Accordingly, the
experimental study of processes with polarized particles becomes more 
and more important. Experiments with deuteron targets or beams
\cite{alexa,kox0,preliminar,cosy_proposal} are particularly
interesting, since
the deuteron serves as a unique source of information on neutron
properties at high transferred momenta; the knowledge of which allows, 
e.g. to check a number of QCD predictions and sum rules.
For example, for an investigation of the $NN$ interaction in the deuteron
at short distances, the three deuteron form factors
(magnetic, electric and quadrupole)   have to be determined.
 In the elastic   $eD$ scattering with unpolarized particles
 one can measure only two independent
 quantities, e.g. the magnetic form factor and   the deuteron
function  $A(Q^2)$, the latter being a kinematical combination
of all three form factors. Even these two quantities reveal
an important information
about the quark physics and dynamics at short 
distances as demonstrated, for instance, in recent
measurements \cite{alexa} at TJNAF. However, for a full determination
of the deuteron form factors separately, one
needs measurements with polarized particles. For instance, measurements of the
tensor analyzing power $T_{20}$ of recoil deuterons in  elastic
 $eD$ scattering allow for a determination of the charge form factor
 $G_c$ at high transferred momenta. Namely the charge form factor
 is very sensitive  to  details of the $NN$
 interaction~\cite{kox0,preliminar,gross} and, besides  information
 about short range correlations in the deuteron, the investigation
 of  $G_c$ may essentially constrain the theoretical models applied
 in this area. However, in spite of the fact that 
the electromagnetic processes are considered as the
 cleanest ones, in such reactions one
 probes mainly the quark structure of the target, leaving almost
 untouched the physics connected with gluon degrees of freedom.
 In this context, hadron deuteron processes can be considered
 as complementary tool in investigating phenomena at short distances and
  also as a source of unique information unavailable in 
electromagnetic reactions,
 such as nucleon resonances, checking non-relativistic effective models,
 $NN$ potentials etc.

 Experimental and theoretical investigations of the proton-deuteron
 processes at intermediate and high energies
 have started some decades ago by studying elastic 
$pD$ scattering~\cite{elsatic},
 exclusive and inclusive break-up reactions \cite{inclusive,cosy} 
with the goal of determining the
 the details of the deuteron wave 
function at short distances and the relevant reaction mechanisms.
 Note, that in elastic backward $pD$ processes it is possible to 
determine completely the reaction amplitude
 by measuring a full set of polarization observables
 (see, e.g., \cite{rekalo,ladygin,ourprc}). Hence,
  as in the electromagnetic processes one 
needs to measure different polarizations of
  the recoil deuteron. 

Since such polarization  observables can be studied
  only by an  additional secondary scattering
  of the reaction products, e.g., inside a polarimeter, it is 
obvious that second process
  must possess a high enough cross section to assure a good  efficiency of
  the polarimeter. Traditionally, at low energies one uses the process
  ${\rm ^3He}\,(D,p)\,{\rm ^4He}$~\cite{grub},
  while for relativistic energies the elastic
  $Dp$ scattering serves as polarimeter \cite{bald}. 
  Bugg and  Wilkin \cite{wil1} proposed to use 
as an efficient deuteron polarimeter
 the process $p{\vec D}\to n(pp)$, where the final  $pp$ pair is
 detected with very low excitation energy (see also ref. \cite{polder}).
It has been argued that at low excitation energies and low transferred momenta
the process $p{\vec D}\to n(pp)$ is determined by the elementary
$pn$ charge-exchange of the incoming proton with the neutron
in the deuteron, whereas the second proton acts merely as a spectator.
In this case, the detected $pp$ pair can be considered to be in the $^1S_0$
final state.  Within the non-relativistic spectator mechanism,  
a significant value of the tensor analyzing
power $T_{20}$ and a vanishing vector analyzing
power were predicted~\cite{wil1}. The corresponding cross section
is rather large, so that the process  $p{\vec D}\to n(pp)$ can be considered
as a good tool for determining the deuteron tensor characteristics. Later
detailed investigations ~\cite{ishida,moto,wil2,kox} of this process 
confirmed the previous theoretical predictions \cite{wil1}.
The charge-exchange
processes of this type have also been proposed for investigations
of other processes with deuterons, e.g.
   $pp\to D\pi^+$ reactions \cite{bugg}, $\Delta N$ systems \cite{kozi},
 $NN\pi$ systems, inelastic  $({\vec D},{\vec D'})$ reactions off
 heavy nuclei  to study isoscalar transitions
  $\Delta T = 0,\, \Delta S = 1$ \cite{morlet} etc.
 In addition, the reaction $p{\vec D}\to n(pp)$ may be 
interesting in investigations
 of the elementary $NN$ amplitude. As shown in ref. \cite{wil1}, 
 this reaction
 can be used as a part of a complete set of experiments to determine completely
the amplitude of charge-exchange reaction $pn\to np$ \cite{glagolev}.
 One may also investigate the influence of the nuclear medium on the $NN$
 amplitude~\cite{heckman,titov}, or to study the double spin flip
 processes in quasi elastic scattering of deuterons from heavy
 nuclei~\cite{ishida}. 

The direct consequence of these facts is that
 nowadays the
 interest in investigations of charge-exchange processes does not abate.
For example, at the cooler synchrotron COSY in FZ J\"ulich 
a program to study
 of similar processes at relativistic energies has
 already started \cite{cosy_proposal,cosy} and a detailed 
investigation of polarization observables
 is envisaged. Inspired by this, in previous work
 \cite{nashyaf} we investigated the
process $p{\vec D}\to n(pp)$ within the impulse approximation.
 The goal of the present  paper is to consider
 the charge-exchange reaction at relativistic energies, as accessible at 
COSY and upgraded Dubna accelerator by (i) taking into account the effects 
of final state interaction, (ii) to check whether  in this 
case the non-relativistic predictions~\cite{wil1}
 hold and (iii) whether the reaction still 
can be regarded as a deuteron polarimeter tool
 and/or as complementary source of an experimental determination
 of the elementary charge-exchange partial amplitudes.
 We propose a covariant generalization of the spectator mechanism~\cite{wil1}
  based  on the Bethe-Salpeter (BS) formalism and on a 
numerical solution of the BS equation
 with a realistic one-boson exchange kernel~\cite{solution,parametrization}.
 Our amplitude of the process in an explicitly covariant form allows
 for a determination of any polarization observables. Nevertheless,
 here we focus 
 on calculations of the cross section and
 the tensor analyzing power $T_{20}$ for kinematical conditions
as relevant for experiments at COSY. Vector analyzing powers of the deuteron
are strictly equal to zero in our case, since we consider the $^1S_0$
NN final state.
The corresponding expressions are quite lengthy, so we do not
 present an explicit comparison with the non-relativistic formulae, 
nevertheless,
 the final results are written in a form as close as possible to the 
non-relativistic case. We are going to compare our results, computed at 
non-relativistic energies, 
with corresponding data and corresponding  non-relativistic
 calculations \cite{wil2} thus demonstrating that 
our formulae hold
 in the non-relativistic limit as well, as it should be. We adopt for the
 process  $p{\vec D}\to n(pp)$ with a slowly
 moving, small-excitation energy $pp$ pair
 the same mechanism as in \cite{wil1}, i.e., the process is treated, in
 the deuteron center of mass, as
 a charge-exchange between the incoming proton and internal
 neutron with the second proton
 as a spectator. The resulting $pp$ pair is
 supposed to be detected solely in the $^1S_0$ state. Particular
 attention is paid to the pure impulse approximation, where the final
 state interaction in the $pp$ pair is, for the time being, disregarded.
 Within the impulse approximation we
 study systematically peculiarities of the reaction to 
find proper kinematical
 conditions where the supposed mechanism is adequate and to
 fix the choice of input parameters. Then
 the final state interaction in the $^1S_0$ state is taken into account
 by solving the inhomogeneous BS equation
 in the one-iteration approximation, which allows one to
numerically compute the corresponding
 partial amplitudes. The effects of final state interaction are found to be
 substantial
 and essentially improve the agreement with data. Methodological prediction
 for the COSY kinematics will be presented as well.

This paper is organized as follow. In Sect. II 
kinematics and notation are introduced.
A short review of
the spin structure of the amplitude
$\displaystyle\frac12 + 1\to\displaystyle\frac12+0$
and definitions of  polarization observables are presented. 
Sect. III deals with the
invariant amplitude within the BS formalism and  defines 
the corresponding BS wave functions.
A reduction of the covariant form of the amplitude to the traditional
form in the two dimensional spinor space is also performed in this section.
In the next section IV a detailed study of relativistic
impulse approximation and a comparison with data  is given.
In  Sect. V the procedure of accounting for  final state interaction
in the continuum within the BS formalism is discussed. The total BS
wave function for the $^1S_0$ configuration is presented, the corresponding
numerical calculations of the cross section and tensor analyzing power
and a comparison with the available experimental data is performed. Conclusions
and summary may be found in Sect. VI. Some cumbersome expressions are
relegated to the Appendices.

\section{Kinematics and notation} \label{gl1}
\begin{figure}[h]    %Fig.1
\centerline{\epsfbox{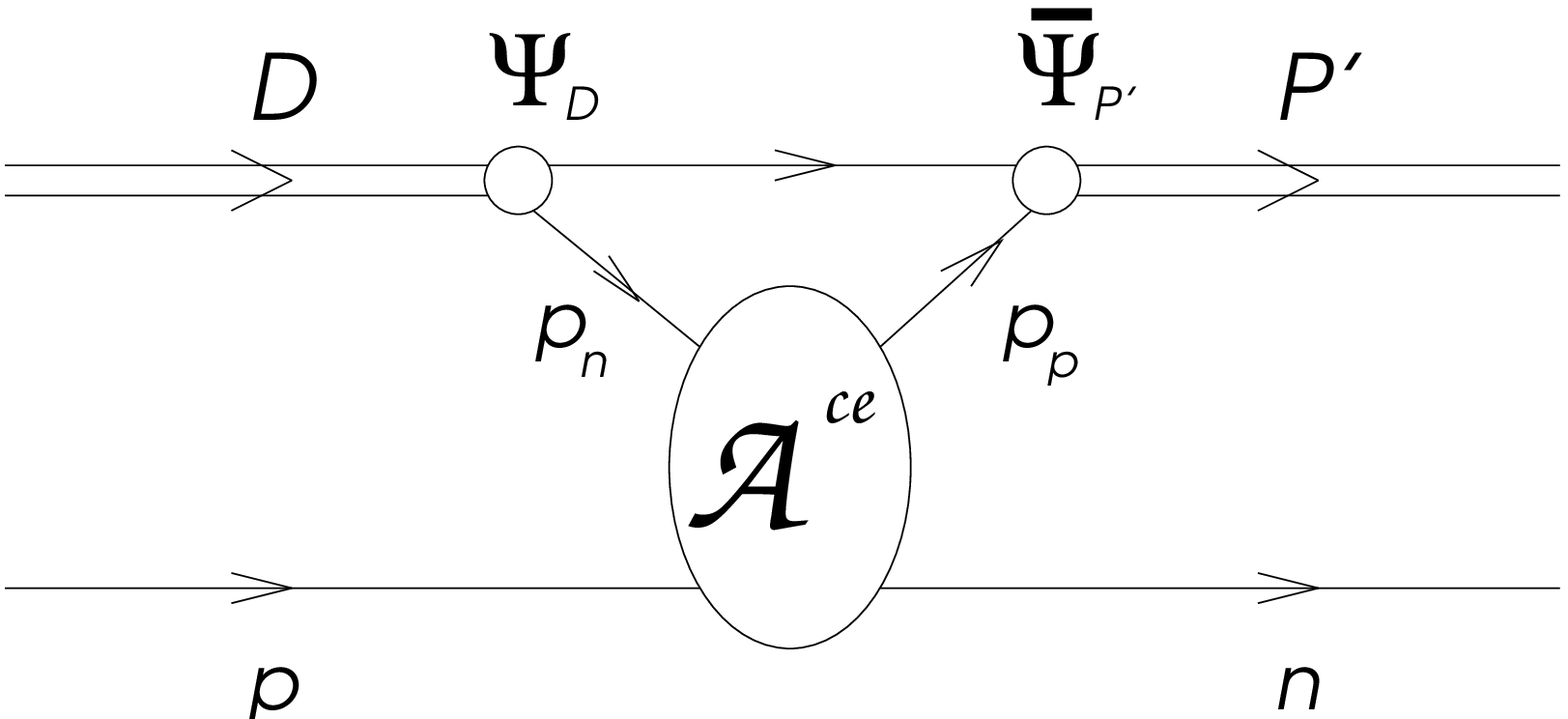}}
\caption{
Spectator mechanism for the  charge-exchange process  $pD\to n(pp)$.
The Bethe-Salpeter amplitude for the deuteron bound
state and the $pp$ pair in the continuum are denoted as $\Psi$ and $\bar \Psi$,
respectively. The elementary $pn$ charge-exchange amplitude is
symbolically represented by ${\cal A}^{ce}$.
}
\label{pict1}
\end{figure}

Since in the exclusive
processes with  three nucleons in the final
state we are interested
in studying  correlations in  the $pp$ pair
we select  those of them which, in
the deuteron center of mass system,
correspond to final states with one fast neutron and
 a slowly moving
proton-proton  pair, i.e. reactions of the type
\begin{equation} p\,+\,\vec D
\,=\, n + (p_1+p_2).  \label{reaction}
\end{equation}
A peculiarity of the processes (\ref{reaction})
is that the transferred momentum from the proton to the neutron
is low, hence the main mechanism of the reaction can be
described as a charge-exchange process of the incoming proton off
the internal  neutron, whereas the second proton in the deuteron remains
merely as a spectator.
As well known (see e.g. ref.  \cite{diu})
the differential cross section of elementary 
charge-exchange process  $pn\to np$
exhibits a sharp maximum at vanishing transferred momenta. Therefore,
if the reaction (\ref{reaction}) is 
indeed governed by a charge-exchange subprocess,
then the resulting  $pp$ pair will be 
detected with low total and relative momenta.
Reactions of this kind can fairly well be distinguished from other 
processes. For relatively low initial energies, such reactions
are quite well experimentally investigated.
 In Fig.  \ref{pict1},  the diagram
 of such  processes is schematically depicted. 
The following notations are adopted:  $p=(E_p,\bp)$ and
$n=(E_n,\bn)$ are the 4-momenta of the incoming proton and outgoing neutron,
 $P'$ is the total 4-momentum of the  $pp$ pair, 
which is a sum of the corresponding
 4-momenta of detected protons, $p_1=(E_1,\bp_1)$, 
$p_2=(E_2,\bp_2)$, $P'=p_1+p_2$.
The invariant mass squared of the pair is
 $s_f,~s_f=P'^2=(2m+E_x)^2$, where  $m$ stands for the nucleon
 mass and  $E_x$ for the excitation energy
 of the pair.
 Conform the supposed reaction mechanism, the excitation energy  $E_x$
   ranges to a few MeV, say $E_x= 0-8$ MeV. At such
   low values of  $E_x$, the main contribution in the final state
 of the $pp$ pair in the continuum comes from the
 $^1S_0$ configuration~\cite{wil2}. In what follows all corrections from
 higher partial waves are neglected, however, we realize that for higher
 values of $E_x$ an increasing role  of these corrections is
  expected.

  Further, the Dirac spinors
\begin{eqnarray}
u(\bp,r)=\sqrt{m+\epsilon} \left ( \begin{array}{c} \chi_r \\
\displaystyle\frac{(\bsigma\cdot\bp)}{m+\epsilon}\chi_r
\end{array}  \right)
\label{spinor}
\end{eqnarray}
 normalized as  $\bar{u}(p) u(p)=2m$ are introduced.
 Then the differential cross section for the reaction (\ref{reaction}) reads
\begin{eqnarray}
d^9\sigma\,=\,\frac{1}{2\sqrt{\lambda(p,D)}}\,|M_{fi}|^2\,(2\pi)^4
\delta (P_f-P_i)\,
\frac{d^3\bn}{2E_n(2\pi)^3}\,\frac{1}{2}
\prod\limits_{k=1}^{2}
\frac{d^3\bp_k}{2E_k(2\pi)^3},
%\,\frac{d^3\bp_2}{2E_2(2\pi)^3}
\label{cross}
\end{eqnarray}
where  $\lambda(p,D)$ is the flux factor, $M_{fi}$ is the invariant amplitude
  and the statistical factor  $1/2$
 is due to two identical particles (protons)
 in the final state.
 By changing in eq. (\ref{cross})
 the kinematical variables from the momenta  $\bp_{1,2}$
 to the relative and total momenta of the pair and taking into
 account that there is no angular dependence in the
 $^1S_0$ configuration the cross section can be written as
\begin{equation}
d^3\sigma\,=\,\frac{1}{16\pi\sqrt{\lambda(p,D)}}\,
\sqrt{1-\frac{4m^2}{s_f}}\,
|M_{fi}|^2\,\frac{d^3\bn}{2E_n(2\pi)^3}\,\frac{1}{2}.
\label{cross1}
\end{equation}
Since our numerical solution of the Bethe-Salpeter equation
has been obtained in the deuteron rest system all further
calculations are performed in that.  The quantization  $z$ axis
is chosen along the momentum  $\bp$
of the incoming proton; the  $x$ and  $y$ axes will be specified below.
Changing the variables in (\ref{cross1}) we arrive at
\begin{equation}
\frac{d^2\sigma}{dt\,ds_f}\,=\,\frac{1}{2}\frac{1}{64\pi\lambda(p,D)}\,
\sqrt{1-\frac{4m^2}{s_f}}\,\int\frac{d\phi}{(2\pi)^3}\,|M_{fi}|^2,
\label{cross2}
\end{equation}
where  $q=n-p$, $t=q^2$, $s_f=(D-q)^2$, and
$\phi$ denotes  the azimuthal angle of the final neutron.
 Further we consider only the case where the initial proton and the final
 neutron are unpolarized  and the polarization density matrix
 of the initial deuteron, $\rho_D$, possesses an axial symmetry relative to the
 $z$-direction, i.e.
\begin{eqnarray}
\rho_D=\frac{1}{3}\,{\bf 1}\,+\,p_v\,{\hat T}_{10}\,+\,p_t\,{\hat T}_{20},
\nonumber
\end{eqnarray}
where  $p_v$ and $p_t$ are the vector and tensor polarization
parameters, respectively. In this case  the angular dependence
upon   $\phi$ in eq. (\ref{cross2}) is trivial. Finally one has\begin{equation}
\frac{d^2\sigma}{dt\,ds_f}\,=\,\frac{1}{2}\frac{1}{64\pi\lambda(p,D)}\,
\sqrt{1-\frac{4m^2}{s_f}}\,\frac{1}{(2\pi)^2}\,|M_{fi}|^2,
\label{cross3}
\end{equation}
where the amplitude  $|M_{fi}|^2$ can be computed at arbitrarily fixed
value of  $\phi$, e.g.,  $\phi=0$.
The procedure of computing the amplitude
$M_{fi}$  consists on several  stages: (i)
we analyze general spin structure
 in terms of a decomposition of  $M_{fi}$ over a relevant
 independent set of spin variables with coefficients
 being  invariant partial spin amplitudes of the process
 since all observables
 can be expressed via these partial amplitudes,
 (ii) the diagram in Fig.~\ref{pict1} is computed explicitly and the
  results for  $M_{fi}$
 are regrouped  to obtain an expression in the same form as for
 the general decomposition of  $M_{fi}$,
 (iii) from the direct comparison of the expression with the
 phenomenological form the partial spin amplitudes are found and
 the polarization observables computed.

By virtue of zero angular momentum of the final pair, the process
(\ref{reaction}) is of the type
$1/2\,+\,1\,=\,1/2\,+\,0$, for which the symmetry restrictions leave only
six independent (complex) partial amplitudes.
The choice of their explicit representation depends upon the kinematical
conditions of the attacked problem. 
One may choose the helicity representation, or
the representation with a given spin projection for specific choices
of the quantization axis etc.
In this paper, we choose the following way to determine the partial
amplitudes (see also  \cite{keaton}):
initial $|i\rangle$ and final  $|f\rangle$ states of the system,
besides other quantum numbers, are characterized by the spin projections
on the $z$ axis; in the matrix element $M_{fi}$
this spin dependence is written explicitly by emphasizing  in
 $|i\rangle$ and  $|f\rangle$ the 3-polarization vector of the deuteron
 and the two-component Pauli spinors for nucleons.
We introduce, in the deuteron center of mass, where  $\bD={\bf 0}$,
 the three basis vectors as follows
\begin{equation}
\bc=\frac{\bp}{|\bp|},\quad
\bb=\frac{[\bp\times\bn]}{|[\bp\times\bn]|},\quad \ba=[\bb\times\bc].
\label{basis}
\end{equation}
 Then the amplitude   $M_{fi}$ can be represented in the form
\begin{equation}
M_{fi}\equiv \CT_{r'r}^M =  [\chi_{r'}^+]_\alpha\,
\left( \CM_{\alpha\beta}\,\bxi_M\right)
[\chi_r]_\beta ,\quad \alpha,\beta=1,2,
\label{ampF}
\end{equation}
where  $r', r$ and  $M$ are the spin projections for the
neutron, proton and deuteron, respectively. The amplitude
  $\CM_{\alpha\beta}$ is a vector in the coordinate space and
  a matrix in the spinor basis and consequently can be
  decomposed over the introduced basis vectors
(\ref{basis}) and Pauli matrices  $\sigma_i$ ($i=x,y,z$)
as
\begin{eqnarray}
\nonumber\CM_{\alpha\beta} & = &
i\, {\cal A}\,  \bb\,\delta_{\alpha\beta}+
{\cal B}\,\bb\,   \,(\bsigma\cdot\bb)_{\alpha\beta}
+ {\cal C}\,\ba\,(\bsigma\cdot\ba)_{\alpha\beta} +\\
& + &
{\cal D}\,\ba \,
(\bsigma\cdot\bc)_{\alpha\beta}+{\cal E}\, \bc \,(
\bsigma\cdot\ba)_{\alpha\beta}+
{\cal F}\,\bc\,  (\bsigma\cdot\bc)_{\alpha\beta}.
\label{amplit}
\end{eqnarray}
In eq.  (\ref{ampF} ), $\bxi_M$ stands for the polarization vector
of the deuteron in its center of mass system:
\begin{equation}
\begin{array}{ccc}
\bxi_{+ 1} =-\displaystyle\frac{1}{\sqrt{2}}\,
\left ( \begin{array}{c} 1 \\ i \\ 0 \end{array} \right ),
&
\quad \bxi_{- 1}=\displaystyle\frac{1}{\sqrt{2}}
\left ( \begin{array}{c}1\\ -i\\0 \end{array} \right ),
&
\quad \bxi_{0}=
\left ( \begin{array}{c} 0\\ 0\\1 \end{array} \right ).
\label{xilab}
\end{array}
\end{equation}
Due to the use of the two-dimensional spin and 3-dimensional
vector representation,  eqs. (\ref{ampF}) and (\ref{amplit})
are not manifestly covariant. Nevertheless, 
such a representation of partial amplitudes
is of most general form and valid in both relativistic and 
non-relativistic
considerations. This may immediately be seen if one  expresses
in covariant matrix
elements the  polarization 4-vector of the deuteron
$\xi_M$ (in any system of reference) via the
3-dimensional  $\bxi_M$ as
\begin{eqnarray} \xi_M=\left [\frac
{(\bD\cdot\mbox{\boldmath{$\xi$}}_M)}{M_D},\,
\mbox{\boldmath{$\xi$}}_M+\bD \frac {(\bD
\cdot\mbox{\boldmath{$\xi$}}_M)}{M_D(E_D+M_D)}\right ],
\nonumber
\end{eqnarray}
and passes   from the  4-spinors defined in eq.
(\ref{spinor}) to Pauli spinors  $\chi_{r}$.

For convenience, the $y$ and $x$ axes are oriented along  $\bb$ and $\ba$,
respectively. In this case  the neutron azimuthal
angle  $\phi$ can be put equal to  zero.
The $z$ axis, as mentioned above, is parallel
to $\bc$. The three variables, upon which the partial amplitudes
 $\CA,\CB,...\CF$ depend,  are chosen to be the total initial
 energy $s$, the transferred momentum
 $t$, and the invariant mass of  the  $pp$ pair $s_f$.
 These amplitudes are related with the spin amplitudes  $\CT_{r'r}^M$
 via the following expressions
\begin{eqnarray}
\begin{array}{ll}
{\cal A}=(\CT_{-\frac{1}{2}-\frac{1}{2}}^1+
\CT_{\frac{1}{2}\frac{1}{2}}^1)/\sqrt{2},
&\quad
{\cal B}=-(\CT_{\frac{1}{2}-\frac{1}{2}}^1-
\CT_{-\frac{1}{2}\frac{1}{2}}^1)/\sqrt{2},
\\
{\cal C}=-(\CT_{\frac{1}{2}-\frac{1}{2}}^1+
\CT_{-\frac{1}{2}\frac{1}{2}}^1)/\sqrt{2},
&\quad
{\cal D}=(\CT_{-\frac{1}{2}-\frac{1}{2}}^1-
\CT_{\frac{1}{2}\frac{1}{2}}^1)/\sqrt{2},
\\
{\cal E}=\CT_{\frac{1}{2}-\frac{1}{2}}^0, 
&\quad
{\cal F}=\CT_{\frac{1}{2}\frac{1}{2}}^0.
\end{array}
\label{eq3}
\end{eqnarray}
Note that having computed
the amplitudes  $\CA,\CB,...\CF$, all
polarization observables for the
process (\ref{reaction}) can be found as  proper combinations
of these partial amplitude.
So, if an operator  $\CO$ corresponds to a  measurable physical
quantity  then its  mean value is
\begin{eqnarray}
\langle \CO\rangle = 6 \frac{\mbox{Tr} \left (\CM \,\CO\,\CM^+\right )}
{\mbox{Tr} \left (\CM \,\CM^+\right ) },
\label{observable}
\end{eqnarray}
where the denominator corresponds to the cross section  of the reaction
(\ref{reaction}) with unpolarized particles
\begin{eqnarray}
 \frac{1}{6} \mbox{Tr} \left (\CM\CM^+\right ) =
\frac{1}{3} (\CA\CA^* + \CB\CB^* + \CC\CC^* + \CD\CD^* + \CE\CE^* + \CF\CF^*).
\nonumber
\end{eqnarray}
For instance, the tensor analyzing power $\langle T_{20} \rangle$
is
\begin{eqnarray}
\nonumber
\langle T_{20}\rangle &=& 6 \frac{\mbox{Tr} \left (\CM
\,{\hat T}_{20}\,\CM^+\right )} {\mbox{Tr} \left (\CM \,\CM^+\right ) }\\
&=& \frac{\sqrt{2}}{\mbox{Tr} \left (\CM \,\CM^+\right ) }\,
(\CA\CA^* + \CB\CB^* + \CC\CC^* + \CD\CD^* - 2\, [\CE\CE^* +
\CF\CF^*]).
\label{t20}
\end{eqnarray}
Note that the representation of  the amplitude  $M_{fi}$
 by eqs. (\ref{ampF}) and (\ref{amplit}) holds if
 initial and final states can be described by wave functions
 (pure spin states), otherwise for mixed  states
 the square of  $M_{fi}$ must be averaged with the spin density matrices
\begin{eqnarray}
\rho_N=\frac12\,\sum_r |r\rangle\langle r|,\quad
\rho_D=\frac13\,\sum_M |M\rangle\langle M|.
\label{dens}
\end{eqnarray}
An explicit covariant  expression
for the deuteron density matrix can be found in ref. \cite{ourphyslett}.
In the present paper the amplitudes  $\CT_{r'r}^M$
and  observables for the reaction   (\ref{reaction})
are evaluated within the BS formalism. It is worth noting that
in theoretical considerations of  high energy reactions
with deuterons as target and two interacting
nucleons in the final state (scattering or bound state)
always, at least one of two-nucleon systems is moving, consequently
  Lorenz boost effects must be treated in a consistent way.
 In our opinion, the most appropriate approach for these purpose
 is the BS formalism where a consistent description
 of the deuteron bound state and scattering states of the $NN$ pair
 as well as the off-mass shellness  of nucleons and Lorenz boost effects
 may be achieved \cite{ourprc}. There are other approaches to the
 relativistic description of  reactions with deuterons. For instance,
 the Gross equation \cite{gross1}, which is a variant
 of the BS approach with one nucleon on-mass shell; it
 provides a covariant description of processes like (\ref{reaction}).
  An analysis of results within the Gross and BS approaches
  shows \cite{ciofi} that for internal relative momenta
   up to $|\bk|\sim 1.5$ GeV/c
  the deuteron amplitudes and wave functions are almost identical.
%  This persuades us that our results and results of formalism of the type
%  \cite{gross1} should be the same.

   In what follows we compute the amplitude  $\CT_{r'r}^M$
 by evaluating the diagram in Fig.~\ref{pict1} within the BS formalism.
Neglecting the initial state interaction between the
 incoming proton and the deuteron and the final state interaction
 of the outgoing neutron with the $pp$ pair simplifies the calculation.
 The initial and final states can then be  written as
  direct products of  spinors of the fast particle and
  the BS amplitudes of the $NN$ system, which correspond to solutions
  of the BS equation for bound (i.e. the initial deuteron) or scattering
  states (i.e. the pair in the continuum).

\section{Invariant amplitude}  \label{gl2}
By using the Mandelstam technique \cite{mandel}
the covariant matrix element corresponding to the diagram in Fig.~\ref{pict1}
 can be written in the form
\begin{eqnarray} \CT_{r'r}^M = {\bar
u}_{\gamma}^{r'}(n)\, u_{\delta}^{r}(p) \int d^4k\,\,
{\bar\Phi}_{P'}(\frac{q}{2}+k)_{\alpha\beta}\,(\frac{\hat D}{2}+{\hat
k}-m)_{\alpha\mu}\Phi_M(k)_{\mu\nu}\,\CA^{ce}_{\beta\gamma,\delta\nu}.
\label{me1}
\end{eqnarray}
In eq. (\ref{me1}) the deuteron
BS amplitude $\Phi_M$ and the conjugate  amplitude ${\bar\Phi}_{P'}$ of
the    $pp$ pair are solutions of the corresponding  BS equation, and
the charge-exchange vertex  $\CA^{ce}$ corresponds to a
 4-point Green function of the subprocess   $pn\to np$
 with, in the most general case, off-mass shell nucleons.

 It is convenient to change from outer products of
 spinors and amplitudes to the usual matrix structures. For
 this sake
 we redefine the BS amplitude \cite{nashi} as
\begin{eqnarray}\nonumber
\Phi(k)\equiv \Psi(k)\, U_C,\quad {\bar\Psi}(k)=
\gamma_0\Psi^\dagger(k)\gamma_0,
\end{eqnarray}
where  $U_C=i\gamma_2\gamma_0$ is the charge conjugation matrix.
Then the new amplitudes $\Psi(k)$ may be considered as usual
  $4\times 4 $ matrices in the spinor space, and the BS equation becomes
an integral  matrix equation. To find a numerical solution of
the BS  equation usually the amplitude is decomposed over a complete
set of matrices, and one solves the
resulting integral equation for the coefficients of such a
decomposition.
These coefficients are known as the partial BS amplitudes.
There are eight independent partial amplitudes for the deuteron, and
the specific form of them depends on the chosen matrix representation.
 In the present paper
we choose the $\rho$-spin representation~\cite{kubis} 
for the partial amplitudes
and, since in the considered reaction the transferred momenta is
rather small, all the amplitudes with at least one negative $\rho$-spin
are disregarded as they may play a role only at high momenta (see,
e.g.  \cite{quad} for the justification). 
Then we are left with two ``++'' partial amplitudes
known as $S$ and $D$ components within the $\rho$-spin classification.
The numerical solution for the BS amplitudes
has been found \cite{solution,parametrization}
by solving the BS equation with a realistic one-boson exchange kernel
including $\pi,\,\omega,\,\rho,\,\sigma,\,\eta,\,\delta$
mesons. In the adopted approximation, the BS amplitude reads (see also
Appendix \ref{sect:aa})
\begin{eqnarray}
\Psi_M(k)=\Psi_{S^{++}}^M(k) +
\Psi_{D^{++}}^M(k).  \label{deu}
\end{eqnarray}

 The charge-exchange vertex  $\CA^{ce}$ is  also  a matrix in the spinor space
 and, for consistency of the approach,
 it would  be preferable to  decompose it  
into partial vertices as in the case of
 the BS amplitude and to find the coefficients from the $NN$ charge-exchange
 reactions. However, this is a rather cumbersome procedure in which one can
 determine only the vertices for on-shell $NN$ processes. 
To extend it to off-shell nucleons
 one needs to specify some method, which inevitably 
requires theoretical models and additional
 approximations. In our case, in the process 
(\ref{reaction}) the transferred momenta and
 energies are considered relatively small, 
hence the virtuality of nucleons in the vertex
 $\CA^{ce}$ may be disregarded. Then  $\CA^{ce}$ 
may be expressed directly via the
amplitude  $f_{r's',sr}$
of the real charge-exchange processes $p+p_n=p_p+n$ 
with all particles on the mass shells
\begin{eqnarray}
f_{r's',sr} =
{\bar u}_{\alpha}^{s'}(p_p)\, {\bar u}_{\beta}^{r'}(n)\,
\CA^{ce}_{\alpha\beta,\gamma\delta}\, u_{\gamma}^{r}(p)\,
u_{\delta}^{s}(p_n).
\label{perez}
\end{eqnarray}
Then
\begin{eqnarray}
\nonumber
\CT_{r'r}^M &=& \sum\limits_{ss'}\frac{1}{(2m)^2}\int d^4k\,
f_{r's',sr}\\
&\times&{\bar u}^{s}(p_n) {\Psi}_M(k)\,(\frac{1}{2}{\hat D}-{\hat k}+m)
{\bar\Psi}_{P'}(k-\frac{1}{2}q)\,u^{s'}(p_p).
\label{fin}
\end{eqnarray}
 Note that  the amplitude  (\ref{fin})
 is manifestly covariant.
For the final  $^1S_0$-state within the  $\rho$-spin classification
the BS amplitude in the center of mass of the $NN$ pair
is represented by four partial amplitudes
$^1S_0^{++}$, $^1S_0^{--}$, $^3P_0^{+-}$ and $^3P_0^{-+}$ \cite{nashi},
which for the sake  of brevity in what follows 
are denoted as $\phi_1,\dots,\phi_4$.
In order to avoid an explicit Lorenz boost  transformation to
the laboratory system, it is convenient to write the  $^1S_0$ amplitude
in a covariant form
\begin{eqnarray}
\sqrt{4\pi}\;{\bar\Psi}_{P'}(p) &=& - b_1\gamma_5 - b_2\frac {1}{m}
(\gamma_5\hat {p}_1 + \hat {p}_2\gamma_5) \nonumber\\
&-&
b_3(\gamma_5\frac{\hat{p}_1-m}{m} - \frac {\hat {p}_2+m}{m}\gamma_5) -
b_4 \frac{\hat{p}_2+m}{m} \gamma_5 \frac{\hat {p}_1-m}{m},
\label{covarj0}
\end{eqnarray}
where  $p_{1,2}=P'/2 \pm p$, and $p$ is the relative momentum. The
four Lorenz invariant functions
$b_i\equiv b_i(P'p,p^2)$ in the center of mass of the pair are
linear combinations of the amplitudes  $\phi_i\equiv
\phi_i(r_0,|\br|),~i=1,\dots,4$ \cite{nashi} (see Appendix \ref{sect:bb}).
Now it is sufficient
to express the amplitude of the process (\ref{reaction}) in terms of
deuteron $"++"$ components and (\ref{covarj0}) to implicitly
account for the Lorenz boost effects \cite{ourprc}.
For  the final state of the pair,
as in the deuteron case,   all the amplitudes
with negative $\rho$-spins
are neglected as well.

Substituting eqs.
(\ref{covarj0}) and  (\ref{deu}) into eq. (\ref{fin}) the matrix element
may be written in terms of two-component spinors and 3-vectors as
\begin{eqnarray}
&& %f_{r's',sr}\;
{\bar u}^{s}(p_n)\, {\Psi}_M(k)\,\left(\frac{\hat P}{2}-{\hat k}+m\right)\,
{\bar\Psi}_{P'}\left(k-\frac{q}{2}\right)\,u^{s'}(p_p)
\nonumber\\ &&
\nonumber
 %f_{r's',sr }
% \hspace*{-1cm}
= \frac{1}{16\pi}\,\frac{m}{E}\,
\frac{\frac{1}{2}M_D-k_0-E}{\sqrt{(E+m)(p_p^0+m)}}
\left \{
\chi^\dagger_s(\bsigma\cdot\bxi_M)\,\chi_{s'}
\left(\psi_{S}-\frac{\psi_{D}}{\sqrt{2}}\right)
\,C_1\right.\\\nonumber
&&+
\chi^\dagger_s(\bsigma\cdot\bk)\,\chi_{s'}(\bk\cdot\bxi_M)\,\psi_{D}\,C_2
 \nonumber\\&&
+\left[-\chi^\dagger_s(\bsigma\cdot\bq)\,\chi_{s'}\,(\bk\cdot\bxi_M)\,
\left(\psi_{S}+\sqrt{2}\psi_{D}\right)+
\chi^\dagger_s(\bsigma\cdot\bk)\,\chi_{s'}\,(\bq\cdot\bxi_M)\,
\left(\psi_{S}-\frac{\psi_{D}}{\sqrt{2}}\right)
\right.
\nonumber\\&&
\left.
\left.
% \hspace*{-1cm}
+\chi^\dagger_s\chi_{s'}\,i\,
([\bq\times\bk]\cdot\bxi_M)\,(\psi_{S}-\frac{\psi_{D}}
{\sqrt{2}})\,\right]\,C_3\,
\right \}\,\phi_1(r_0,|{\bf r}|)) ,
\label{mel}
\end{eqnarray}
where the  quantities  $C_1, C_2, C_3$ have a pure kinematical origin and
are independent of spin variables. Their explicit form
can be found in the Appendix \ref{sect:bb}.
Now, from eq.  (\ref{mel}) it is clearly seen how to compute  $\CT_{r'r}^M$
at given spin variables  $r', r, M$ and, consequently,
the invariant amplitudes
$\CA,\CB,...\CF$ (\ref{eq3}) and observables  (\ref{observable}), (\ref{t20}).
The partial amplitudes  $\phi_i$ may be found from
the BS equation, which, in the simplest case of
pseudo-scalar exchanges reads as
\begin{equation}
\label{neodn}
{\bar\Psi}_{P'}(p)={\bar\Psi}_{P'}^0(p)+ig^2_{\pi
NN}\int\frac{d^4p'}{(2\pi)^4}\,
\Delta(p-p'){\tilde S}(p_2)\gamma_5{\bar\Psi}_{P'}(p')\gamma_5 S(p_1),
\end{equation}
where  $\Delta$ and  $S$ are the scalar and spinor
propagators respectively,
${\tilde S}\equiv U_C\,S\,U_C^{-1}$, and  ${\bar\Psi}_{P'}^0(p)$ is the
free amplitude corresponding to two non-interacting nucleons
(the relativistic plane wave). The solution of eq. (\ref{neodn})
may be presented as a Neumann-like series, the first term of which is the
first one from eq. (\ref{neodn}):
\begin{equation}
\label{rash}
{\bar\Psi}_{P'}(p)={\bar\Psi}_{P'}^0(p)+{\bar\Psi}_{P'}^i(p).
\end{equation}
The second part in eq. (\ref{rash}) is entirely determined by
the interaction and may be symbolically referred  to as scattered wave.
For the  $^1S_0$ state one has
\begin{eqnarray}\nonumber
{\bar\Psi}_{P'}^0(r)|_{P'=(\sqrt{s_f},{\bf
0})}&=&\phi_1^0(r_0,|\br|)\,\Gamma_{^1S_0^{++}}({\hat \br}),\\\nonumber
\phi_1^0(r_0,|\br|)&=&2\,(2\pi)^4\frac{1}{\sqrt{4\pi}}\,\frac{1}{|\br^*|^2}
\,\delta(r_0)\,\delta(|\br|-|\br^*|),
\end{eqnarray}
where  $r=(r_0,\br)$ is the relative 4-momentum of the pair (the
variable of the momentum space),
$|\br^*|=\sqrt{s_f/4-m^2}$ is the  experimentally measured
relative 3-momentum of the pair,  $\Gamma_{^1S_0^{++}}({\hat \br})$ is
the spin-angular harmonic for the $^1S_0$ state \cite{nashi}.
%\ref{sec:app1}.
To determine the scattered wave in eq. (\ref{rash})
it is necessary to solve the BS equation of the type
(\ref{neodn}) including all the above mentioned exchange mesons. 
Solving the BS equation in the continuum
is a much more  cumbersome procedure than solving the homogeneous BS equation.
Besides difficulties encountered in solving the latter (singularities of
amplitudes, poles in propagators, cuts etc.) the former even does not allow
the usual Wick rotation  \cite{wick} to the Euclidean space,
and there are no rigorous mathematical
methods to solve eq. (\ref{neodn}) in the Minkowsky space
\footnote{Actually there is one  realistic
solution of the inhomogeneous BS equation in the ladder approximation,
obtained by Tjon~\cite{Tjonsol}.}. 
However, an approximate solution of
eq. (\ref{neodn}) may be obtained by employing the so-called
"one-iteration approximation" \cite{ourprc}, within which one may obtain
a rather good estimate of the interaction term (see below).

\section{Relativistic impulse approximation}
We start our analysis of the reaction  (\ref{reaction}) by disregarding
the interaction term in eq. (\ref{rash}), i.e. putting
\begin{equation}
\label{appr}
\phi_1(r_0,|\br|)=\phi_1^0(r_0,|\br|),\quad \phi_2=0,\quad
\phi_3=0,\quad \phi_4=0.
\end{equation}
 In this case the final state
of the $pp$ pair is described by the free part $\phi_1^0(r_0,|\br|)$,
what  obviously means  the $^1S_0$ part
of two plane waves. Within the impulse approximation all formulae become
much simpler and one may preliminary investigate the main features of the
process (\ref{reaction}), fix parameterizations 
of the elementary charge-exchange
amplitude, find  proper kinematical regions where 
the  assumptions made hold etc.

%For the current kinematical configuration (see eq. (\ref{fin}))
One has:
\begin{eqnarray}
\delta(r_0)&=&\delta[(P',k-\frac{1}{2}q)/\sqrt{s_f}],\\
\delta(|\br|-|\br^*|)&=&\delta\left(
\sqrt{-(k-\frac{1}{2}q)^2}-\sqrt{\frac{1}{4}s_f-m^2}\right),
\end{eqnarray}
and
\begin{eqnarray}
\nonumber\phi_1^0(r_0,|\br|)&=&2\,(2\pi)^4\,
\frac{1}{\sqrt{4\pi}}\,\frac{1}{E|\bk||\bq|}
\sqrt{\frac{s_f}{\frac{1}{4}s_f-m^2}}\\
&\times&\delta(k_0-[\frac{1}{2}M_D-E])\,\delta \left( \mbox{cos}\,\theta_{kq}
+\frac{s-2E(M_D-q_0)}{2|\bk||\bq|}\right).
\label{plos}
\end{eqnarray}
In eq. (\ref{plos}),
 $\theta_{kq}$ is the angle between  $\bk$ and $\bq$. Then, with  eqs.
  (\ref{appr}), (\ref{plos}) the matrix element in
(\ref{fin}) reads
\begin{eqnarray}
\nonumber
\CT_{r'r}^M &=&
\sum\limits_{s,s'}
\int\limits_{|{\bk}|_{\rm min}}^{|{\bk}|_{\rm max}}
\sqrt{\frac{M_d}{2\pi}}\,
\frac{|\bk| d|\bk|}{|\bq|E\sqrt{(\frac{1}{4}s_f-m^2)(E+m)(p_p^0+m)}}
\;\int\limits_0^{2\pi}d\phi_k\,f_{r's',sr}\,\\
%\nonumber
&\times&\left\{
\chi^\dagger_s(\bsigma\cdot\bxi_M)\chi_{s'}
\left( U_S-\frac{U_D}{\sqrt{2}}\right )
\left( \frac{1}{2}s_f+m P_0'\right )\right .\nonumber\\
&+&
\chi^\dagger_s(\bsigma\cdot
R_q\bk)\chi_{s'}(\bxi_M\cdot R_q\bk)\frac{3}{\sqrt{2}} U_D\frac{P_0'}{E-m}
\nonumber\\ &-&
\chi^\dagger_s(\bsigma\cdot\bq)\chi_{s'}
(\bxi_M\cdot R_q\bk)\left( U_S+\sqrt{2}U_D \right)+
\chi^\dagger_s(\bsigma\cdot
R_q\bk)\chi_{s'}(\bq\cdot\bxi_M)\left(U_S-\frac{U_D}{\sqrt{2}}\right)
\nonumber \\
&+&\left.\delta_{ss'}i([\bq\times R_q\bk]\cdot\bxi_M)
\left(U_S-\frac{U_D}{\sqrt{2}}\right) \right\}.
\label{fin2}
\end{eqnarray}
Here we introduced the corresponding
BS wave functions
\begin{equation}
U_{S,D}\equiv\displaystyle\frac{G_{S,D}}{4\pi\sqrt{2M_d}(M_d-2E)},
\label{functions}
\end{equation}
where $G_{S,D}$ are the BS vertices \cite{quad}.
In this notation,   the  introduced quantities $U_{S,D}$
correspond fully with the non-relativistic $S$ and $D$
wave functions of the deuteron, and the
non-relativistic treatment of the results in terms of usual wave functions
becomes more transparent.
In eq. (\ref{fin2}), the matrix   $R_q$ describes the rotation about the
 $y$ axis by an angle $\theta$ (the angle 
between  $\bq$ and  $z$ axis) defined as
\begin{eqnarray}\nonumber
R_q=\left(\begin{array}{ccc}
\mbox{cos}\,\theta&0&\mbox{sin}\,\theta\\
0&1&0\\-\mbox{sin}\,\theta&0&\mbox{cos}\,\theta
\end{array}\right),
\end{eqnarray}
and the limits of integration over  $|\bk|$
are as follows
\begin{eqnarray}
\nonumber
|{\bk}|_{\rm max, min}=\Biggl|\sqrt{1+\frac{\bq^2}{s_f}}\;
\sqrt{\frac{1}{4}s_f-m^2}\;\pm\;
\frac{1}{2}|\bq|\Biggr|.
\end{eqnarray}
Now it is straightforward to compute   $\CT_{r'r}^M$
for any value of  spin indices  and, by virtue with eq. (\ref{eq3}), 
the quantities
 $\CA,\CB,...\CF$.
 Here it is worth commenting how the charge-exchange amplitude  $f_{r's',sr}$
 is involved into our numerical calculations.
 Beside spin indices this amplitude also depends upon two Mandelstam
 invariants being  the total energy of the subprocess of charge-exchange,
 $s_{pn}=(D/2+k+p)^2$ (where $p$ is the momentum of the initial proton),
 and the invariant transferred momentum  $t=(n-p)^2$, 
which is a common variable for
 both the full reaction (\ref{reaction}) and the subprocess of
 $NN$ reaction. From kinematics, the 4-momenta of the "initial"
 neutron and "final" proton are   $D/2+k$ and  $P'/2+k-q/2$, respectively,
 which might be both off-mass shell.
 In the relativistic impulse approximation
 after integration with the   $\delta$ function, the final proton
 receives an on-mass shell momentum,  $p_p=P'/2+k-q/2$. Consequently here
 only the neutron from the deuteron remains off-mass shell. 
Hence we are left with a
 charge-exchange amplitude with one  nucleon off-mass shell and
 a varying $\sqrt{s}$. In ref. \cite{diu}, the dependence of the 
charge-exchange
 amplitude upon the initial energy has been found to be rather weak (at initial
 energies in a range of few GeV).  Therefore in our calculations, by
   neglecting the off-mass shellness
 of the neutron,   the charge-exchange amplitude
 is taken from the real $NN$ process at  equivalent values of
 $\sqrt{s}$
 (the invariant energy  of the incoming proton and internal neutron).
 Moreover, since this  amplitude is essentially independent  of
 the azimuthal  angle
 $\phi_k$, it is   taken out from the
 corresponding integration.
Such a  procedure of implementation of real amplitudes
into calculations where one or several nucleons are off the mass shell
 is commonly adopted in  literature 
\cite{wil1,kolybasov,ourfewbody,ourphyslett}
 with an "a posteriori" justification from the comparison of
 numerical results with experimental data.
\subsection{Numerical results} \label{gl3}
In our numerical calculations we employ a
parameterization of the elementary charge-exchange amplitude from
ref.~\cite{diu} and parameterizations from partial-wave
analysis performed by different groups~\cite{swart,saidpap}, which are 
available
via  web, cf. \cite{www,said}.
The partial charge-exchange amplitudes are available in the helicity
basis as partial helicity amplitudes
\begin{eqnarray}
&&\nonumber
f_1=\langle\,++\,|\,\CA^{ce}\,|++\,\rangle,\quad
f_2=\langle\,++\,|\,\CA^{ce}\,|--\,\rangle,\quad\\
&&f_3=\langle\,+-\,|\,\CA^{ce}\,|+-\,\rangle,\quad \nonumber\\&&
f_4=\langle\,+-\,|\,\CA^{ce}\,|-+\,\rangle,\quad
f_5=\langle\,++\,|\,\CA^{ce}\,|+-\rangle,\quad \label{helicity}
\end{eqnarray}
 normalized as
\begin{equation}
\frac{d\sigma^{ce}}{dt}=\frac{1}{32\pi\,s(s-4m^2)}
\left\{\sum\limits_{i=1}^4\,|f_i|^2+4|f_5|^2\right\}
\label{chexhnn}.
 \end{equation}
 Since in our matrix element
 (\ref{fin2}) the spin amplitudes $f_{r's',sr}$ are defined in the deuteron
 center of mass,  the helicity amplitudes
 (\ref{helicity})
 must be first boosted along
 the direction  $\bp+\bk$ from the center of mass of the 
 $pn$ system to the laboratory system, and then 
   transformed to spin amplitudes by Wigner rotations.
 Taking into account that the procedure of
 Lorenz boost itself for each
 particle results in an additional  helicity Wick
 rotation~\cite{bourrely}, one needs eight rotations for each
 amplitude in eq. (\ref{helicity}) (see Appendix \ref{sect:cc}  
for details).

In the Fig.  \ref{pict2}
the partial helicity amplitudes (\ref{helicity})
are presented as a function of the transferred momentum
 $|\bq|$ and energy corresponding to the initial momentum
 in the laboratory system  $|\bp|=2.5$ GeV/c.
 The full lines represent the partial-wave analysis~\cite{saidpap}
 whereas the dashed line depict results of the analytical parameterization
  from ref.~\cite{diu}.
  Both parameterizations describe equally well the unpolarized
  charge-exchange cross section,
  nevertheless a substantial
  difference  is seen between two sets of parameterizations.
 Obviously, the unpolarized cross section is not sensitive to
 details of partial amplitudes, and  for a more precise determination
 one needs more independent measurements of polarization observables.
In this context, we remark that such reactions
can be considered as an additional source of information
about the elementary amplitude, as pointed out, e.g., in ref. \cite{glagolev},
since the  process (\ref{reaction}) is entirely
governed by the elementary subprocess of $NN$ charge-exchange.

In what follows we are interested in systematical
calculations of  the cross
section and  tensor analyzing power of the process (\ref{reaction})
for  kinematical conditions achievable at 
COSY~\cite{cosy_proposal}. However, first
we perform calculations for such kinematical conditions for which
experimental data are already available~\cite{kox}. Note that experimentally
one measures the cross section averaged over some interval of the excitation
energy of the $pp$ pair. Conformly, we define the differential cross
section $d\sigma/dt$ as the  double differential cross
section (\ref{cross3}) averaged
 over a given bin of energy
\begin{eqnarray}
\left(\frac{d\sigma}{dt}\right)_k\,=\,
\frac{1}{(8\pi)^3\lambda}\,
\int\limits_{R_k}\,ds_f\,\sqrt{1-\frac{4m^2}{s_f}}\,|M_{fi}|^2,\quad
k=1,2,3...,
\label{sech}
\end{eqnarray}
where  $k$ labels the 
intervals of $E_x$ given in the experiment. 
At  SATURN-II~\cite{kox}, where  the
process  (\ref{reaction}) has been investigated in details at
initial momenta of protons
 $|\bp|= 0.444$ and  $0.599$ GeV/c,
 the mentioned intervals of  $E_x$  are
\begin{eqnarray}
&&R_1:\quad 0\leq E_x \leq 1~MeV,\label{int1}\\
&&R_2:\quad 1\leq E_x \leq 4~MeV,\\
&&R_3:\quad 4\leq E_x \leq 8~MeV.
\label{int3}
\end{eqnarray}
The intervals $R_1$ and $R_2$ fit into the COSY kinematics~\cite{cosy_proposal}
 as well.
Note, that under the kinematical conditions (\ref{int1}-\ref{int3})
the variable $t$ is indeed small,
ranging  in an interval from $ 0$  to  $0.16$  (GeV/c)$^2$.

\begin{figure}[hb]   %Fig3.
\centerline{\epsfbox{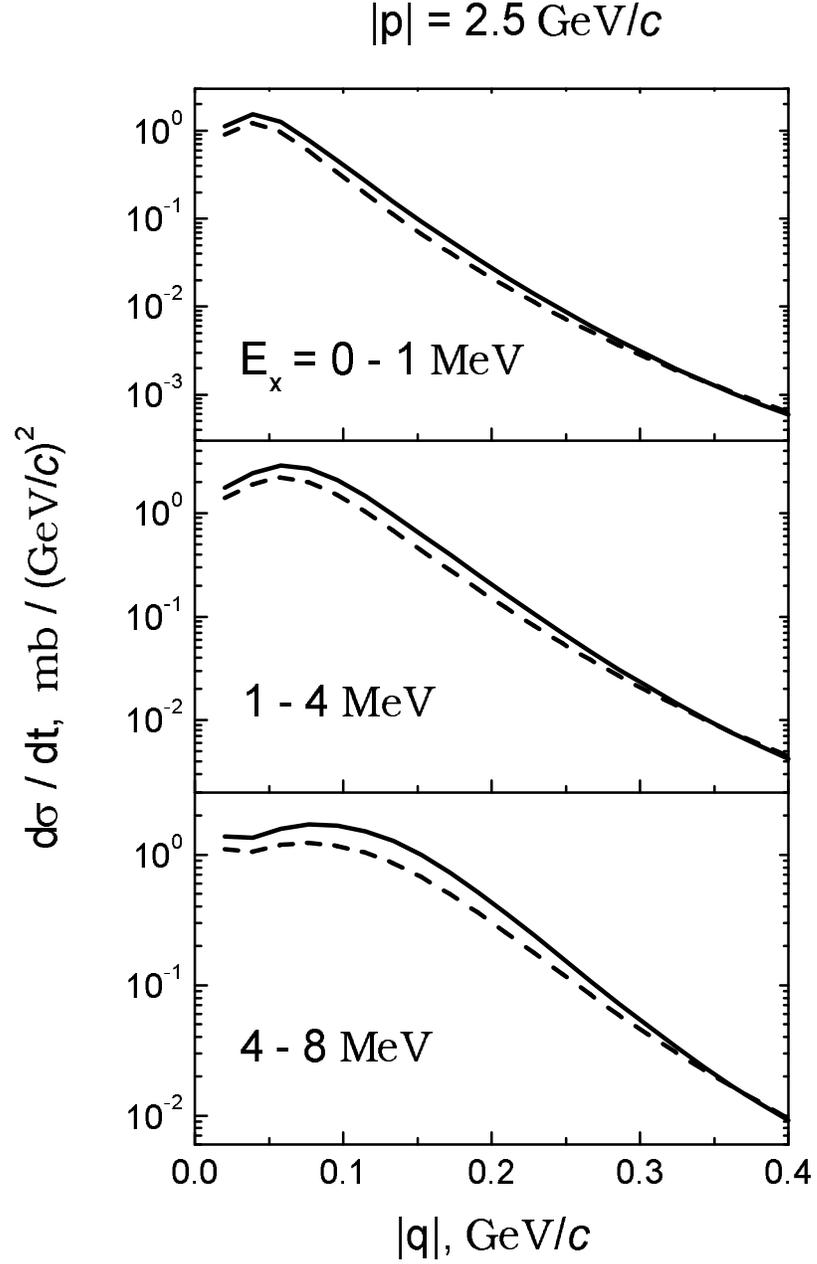}}
\vfill
\caption{Unpolarized cross section, eq. (\ref{sech}),
calculated within the relativistic impulse approximation at a typical
COSY initial momentum for several bins 
of excitation energy of the $pp$ pair. Results of calculations  with two
parameterizations of the $NN$ charge-exchange amplitude  are exhibited
(solid curves are
obtained with the amplitude from \cite{said,saidpap},
the dashed curves from  \cite{diu}).
}
\label{pict3}
\end{figure}

\begin{figure}[hb]   %Fig4.
\centerline{\epsfbox{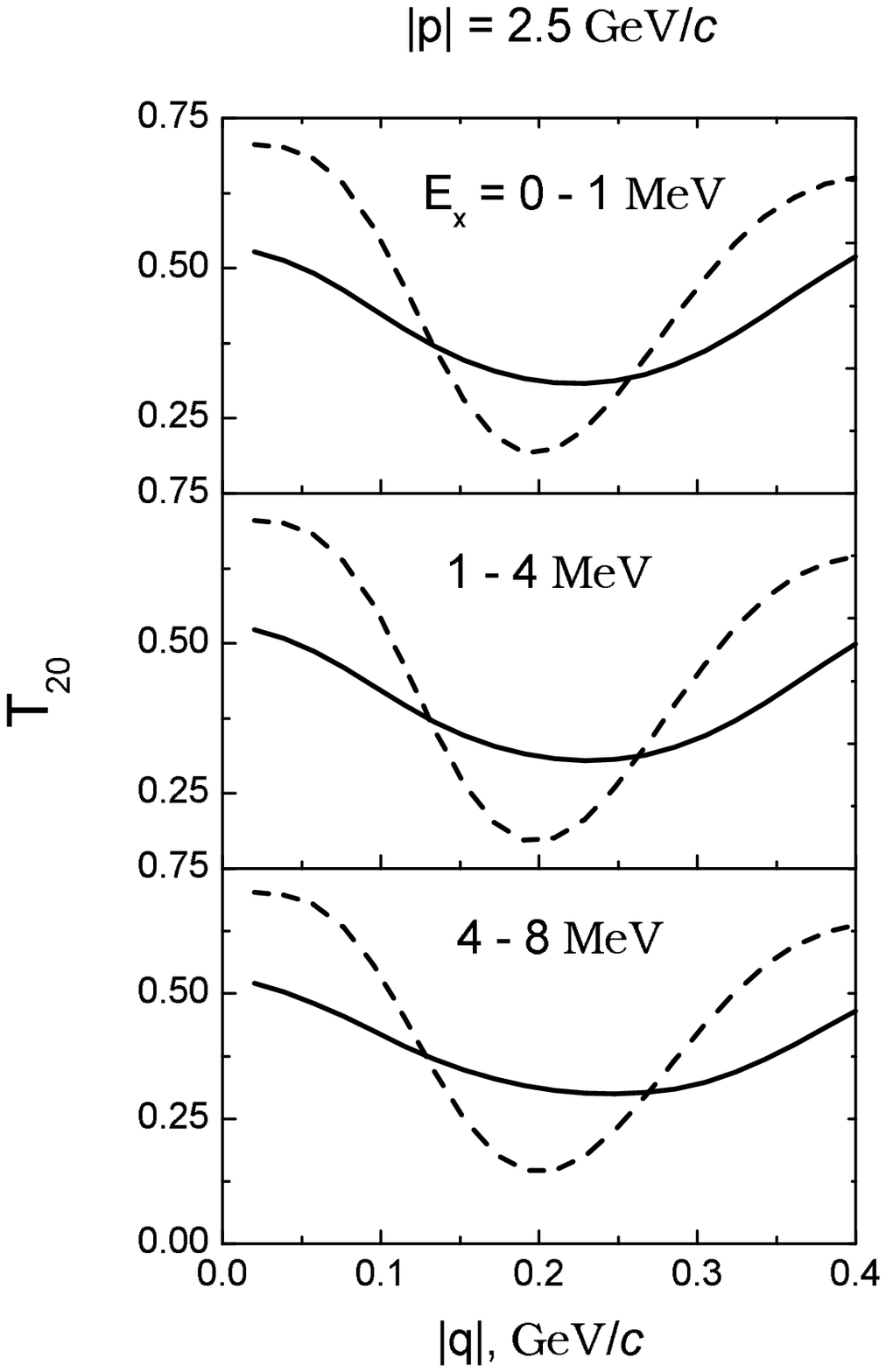}}
\vfill
\caption{The same as in Fig.~\ref{pict3}
but for the tensor analyzing power  $T_{20}$ (\ref{t20}).
}
\label{pict4}
\end{figure}

In Figs. \ref{pict3} and  \ref{pict4} the cross section
 and tensor analyzing power  $T_{20}$
 for the process
(\ref{reaction}) are presented. The initial energy  corresponds to
a typical COSY momentum $|\bp|=2.5$ GeV/c.
The solid lines depict results with the elementary amplitude
taken from ref.  \cite{saidpap}, whereas the dashed lines are results
with the parameterization from ref.~\cite{diu}.
As expected, since different parameterizations equally
well reproduce the elementary charge-exchange cross section,
the unpolarized  cross section of the process (\ref{reaction}) is
not sensitive  to
 parameterizations of the  elementary amplitude.
An opposite situation occurs when calculating the polarization
observables defined in eq. (\ref{observable})
for which the contribution of partial 
amplitudes is non-diagonal and interferences
might be important.  This is clearly illustrated in  Fig. \ref{pict4},
where the tensor analyzing power  (\ref{t20}) exhibits indeed 
a strong sensitivity
to parameterizations of partial amplitudes and practically
does not depend upon the chosen bin of the excitation energy.
From this picture one may conclude that an experimental investigation of the
tensor analyzing power may constrain further  parameterizations of the
  elementary charge-exchange amplitude at high energies.

\begin{figure}[htb]   %Fig.5
\epsfxsize 5in
\centerline{\epsfbox{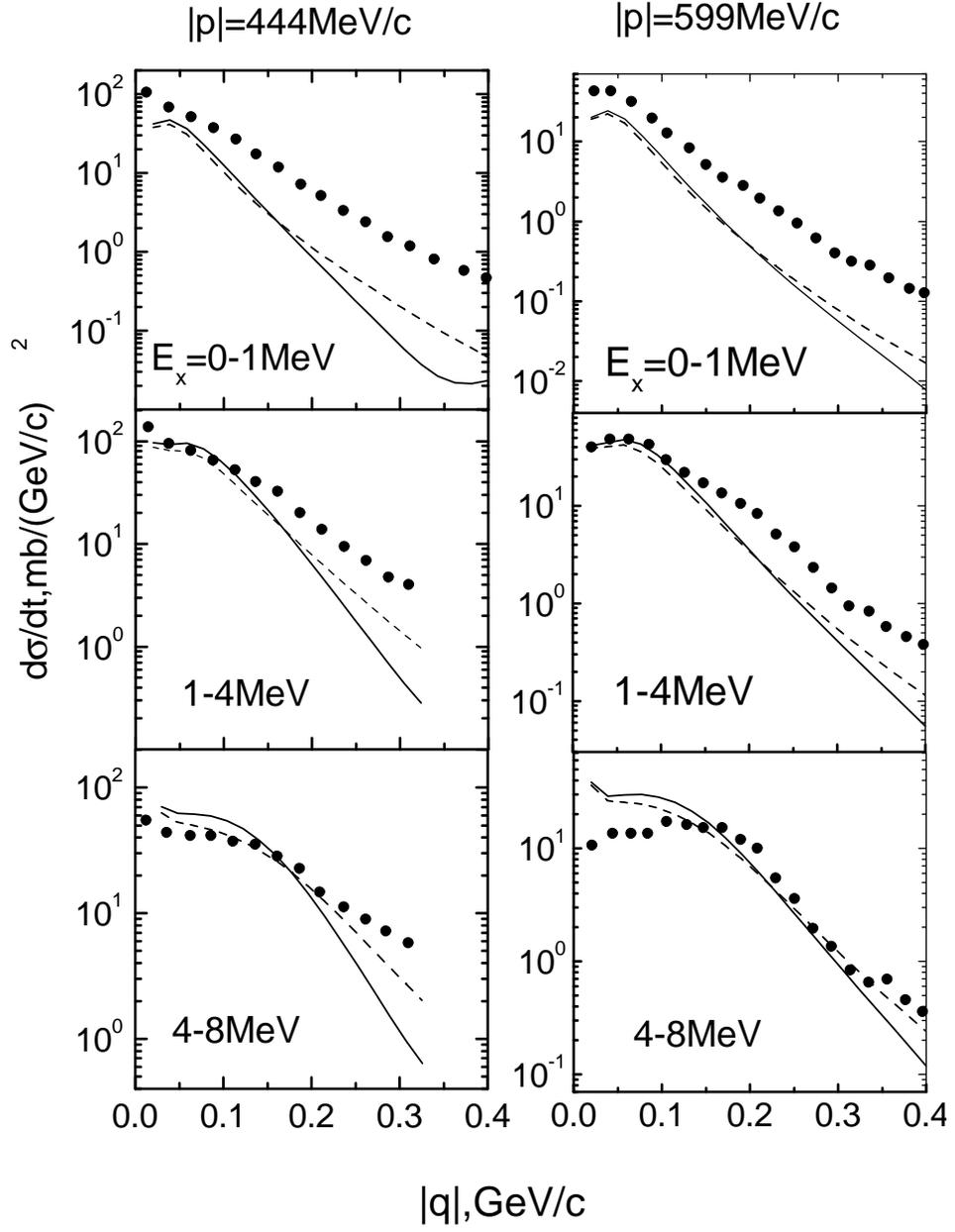}}
\vfill
\caption{Comparison of the  
non-polarized differential cross section eq. (\ref{sech})
within the impulse approximation at non-relativistic
initial energies  ($|\bp|=0.444$ GeV/c
(left panel) and  $|\bp|=0.599$ GeV/c (right panel)), with
experimental
data \cite{kox}. 
Solid curves correspond to the elementary charge-exchange amplitude
from the partial wave analysis from refs.~\cite{said,saidpap},
dashed curves use  \cite{diu}.
}
\label{pict5}
\end{figure}

\begin{figure}[htb]  %Fig.6
\epsfxsize 5in
\centerline{\hspace*{-1cm} \epsfbox{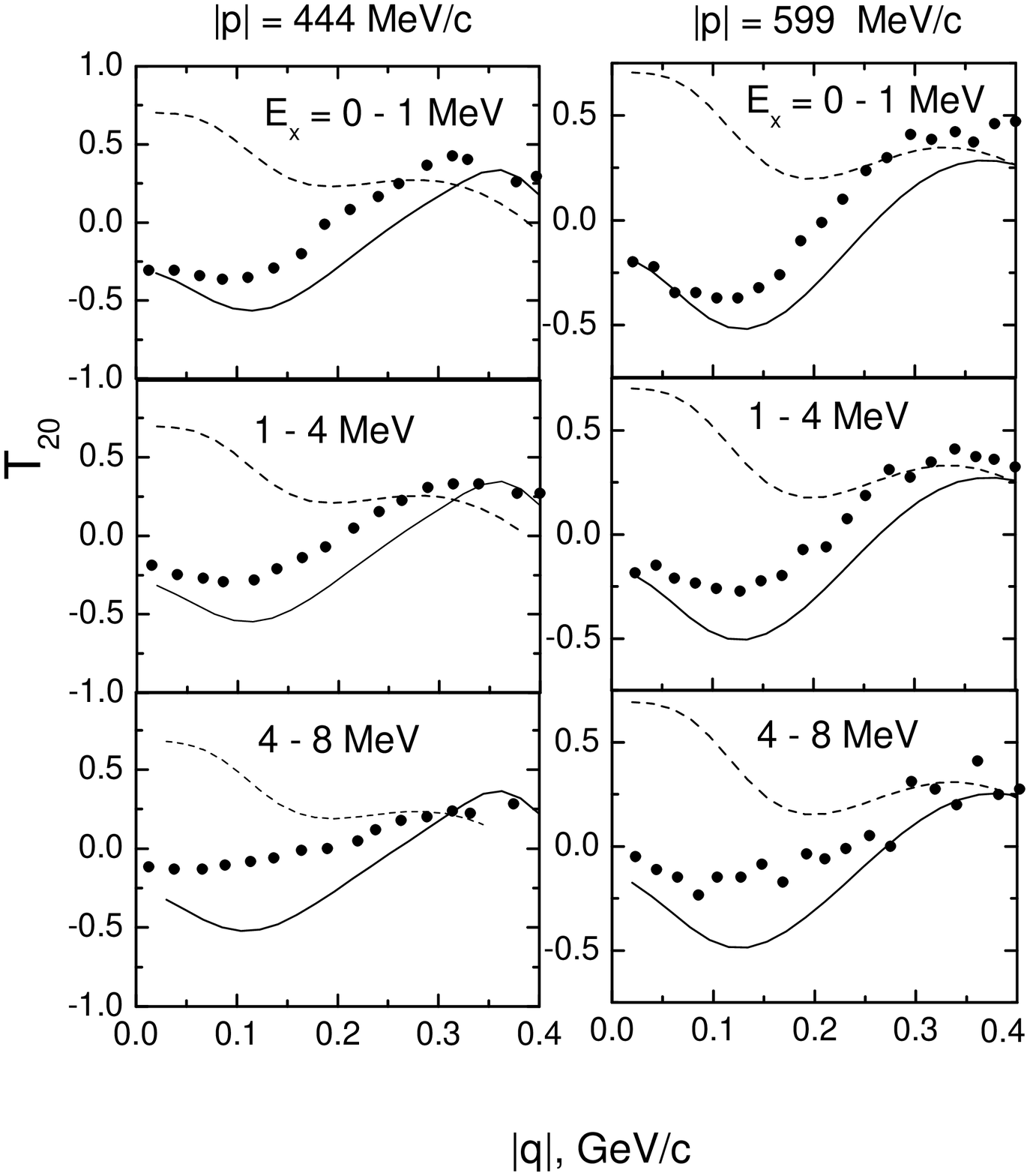}}
\vfill
\caption{The  same as in Fig. \ref{pict5} but for the
tensor analyzing power $T_{20}$ (\ref{t20}).
}
\label{pict6}
\end{figure}

As already mentioned,
 the  process  (\ref{reaction})
 has been experimentally investigated at
 SATURN-II~\cite{kox}. 
  Although   at such energies the final state interaction cannot be
  neglected  and the simple impulse approximation is too rough,
  a comparison of data with theoretical results is rather instructive.
In Figs.   \ref{pict5} and  \ref{pict6} we present
results of calculations of the cross section   
(\ref{sech}) and tensor analyzing
power  $T_{20}$ (\ref{t20}) defined by eqs. (\ref{fin2}-\ref{sech})
together with available experimental data. 
The full lines have been obtained with
parameterizations of the elementary amplitude from refs.
\cite{swart,saidpap}, while the dashed lines with the parameterization
given in ref. \cite{diu}.
From Fig. \ref{pict5} it is seen that the impulse approximation
qualitatively describes the general shape of the cross section as a function
of the transferred momentum $|\bq|$.
For low values of $|\bq|$, say up to  0.2 GeV/c, and in the interval
of the pair excitation energy
$1\leq E_x\leq 4$ MeV here is 
even a good agreement with data, in contrast with
other intervals of   $E_x$ and higher values of  $|\bq|$.
From this and from the results of non-relativistic calculations~\cite{wil1},
where final state interaction and higher partial waves have been taken
into account, one may conclude that at high values of transferred momentum
the effects of final state interaction in the $^1S_0$ state become
dominant. At higher excitation energies the interaction effects are not so
significant, however here corrections from other partial waves
may become important. 
The same conclusions can be drawn from Fig.~\ref{pict6},
where the tensor analyzing power, computed with two parameterizations
(as above, the solid lines correspond
to ref. \cite{swart,saidpap}, dashed curves to ref.  \cite{diu}),
is compared with experimental data. From Fig.~\ref{pict6} it is also
obvious that a qualitative agreement with data for the tensor analyzing power
may be achieved only by using the elementary charge-exchange amplitude
from the partial-wave analysis~\cite{swart,saidpap}, while the parameterization
\cite{diu} results even in opposite sign for  $T_{20}$. This is a direct
indication that a more sophisticated partial-wave analysis gives
more reliable partial helicity amplitudes. Nevertheless, since such an analysis
has been performed for low and intermediate  energies (up to few GeV), a
  further tuning of partial amplitudes (\ref{helicity}) at relativistic
  energies is still desirable. Together with the proper choice of kinematics
  at high energies
  (i.e. a kinematical situation 
where the role of higher partial waves in the final state,
  e.g., triplet states, is suppressed  \cite{glagolev})
  one may expect that the proposed mechanism will adequately describe reactions
  of the type (\ref{reaction}) and corresponding information may be obtained.
   Note that in all
  the above calculations the vector analyzing power is strictly zero.

\section{One-iteration approximation}

 As mentioned, for a consistent
 relativistic analysis of reactions with deuterons and two interacting
 nucleons in the continuum
 one should solve the BS equation for both bound state and scattering
 state within the same interaction kernel. We have found a numerical
 solution for the deuteron bound state with a realistic one-boson exchange
 potential \cite{solution}. The BS equation, after a partial
 decomposition over a complete set of matrices in the spinor space,
 has been solved numerically by using 
an iteration method. We found that the iteration
 procedure converges rather quickly 
if the trial function is properly chosen, e.g.,
 if in the BS equation the combination of the type (\ref{functions}) is
  used as trial functions with 
non-relativistic solutions of the Schr\"odinger equation.
 In such a case, even after the first 
iteration, the BS solution coincides with the exact
 one up to relative momentum $p\sim 0.6-0.7$ GeV/c. 
This circumstance can be used
 if one needs an approximate solution 
of the BS equation at not too large momenta
 $p \leq 0.5-0.7$ GeV/c. This is just 
our case, since in reaction (\ref{reaction})
 the relative momentum of the $pp$ pair is expected to be rather small and
 the scattering part of the amplitude (\ref{rash}) can be obtained from
 eq. (\ref{neodn}) by one iteration, provided the trial function
 is properly chosen.

\subsection{Formalities}
 To solve eq. (\ref{neodn}) we proceed as follow (cf.
 ref. \cite{ourprc}): (i) for simplicity, in the inhomogeneous BS equation  we
 leave only the pseudo-scalar isovector 
exchange ($\pi$-mesons), (ii) write instead
 of eq. (\ref{neodn}) the mixed BS equation by 
introducing in both the left hand side and
the free term the BS vertices, i.e.
\begin{eqnarray}
 G_P(p) = G^0_P(p)-ig^2_{\pi NN}\int\frac{d^4p'}{(2\pi)^4}
 \frac{\gamma_5 \Psi_P(p') \gamma_5}{(p-p')^2-\mu_\pi^2},
 \label{mixed}
 \end{eqnarray}
(iii) bearing in mind that the BS partial vertices may be obtained
from the same spin-angular functions $\Gamma_\alpha({\bf p})$, by replacing
${\bf p}\leftrightarrow-{\bf p}$~\cite{quad},
 we write the corresponding partial BS equation
 \begin{eqnarray}
\nonumber
G_\1s0(p_0,|{\bf p}|) &=& G_\1s0^0(p_0,|{\bf p}|) \\
&-&
ig^2_{\pi NN}\int\frac{d^4p'd\Omega_p}{(2\pi)^4}
\frac{(E_pE_{p'}-m^2+({\bf p\, p}'))}{E_pE_{p'}}
\frac{\phi_1(p_0',|{\bf p}'|)}{(p-p')^2-\mu_\pi^2}.
\label{one_iter_part}
\end{eqnarray}
Further by disregarding the dependence  upon $p_0$  in the meson propagator
 in eq. (\ref{one_iter_part}) and then using the standard  representation
 of propagators via generalized Legendre polynomials $Q_l$ and
 restoring the BS equation in terms of partial amplitudes, one obtains
\begin{eqnarray}
\nonumber
&&
\phi_\1s0(p_0,|\bp|)=\phi_\1s0^0(p_0,|\bp|)-\frac{g_{\pi NN}^2}{4\pi}
\frac{1}
{\left(\frac{1}{2}\sqsf-E_p\right)^2-p_0^2}\times\\
&&\int\limits_0^\infty\frac{d|\bp'|}{2\pi}
\,\frac{|\bp'|}{|\bp|}\,\frac{1}{E_pE_{p'}}\left[(E_pE_{p'}-m^2)
Q_0({\tilde y}_\mu)-|\bp||\bp'|Q_1({\tilde y}_\mu)\right]
 u_{^1\!S_0}(s_f,|\bp'|),
\label{before_int}
\end{eqnarray}
where
${\tilde y}_\mu=\displaystyle
\frac{\bp^2+\bp^{'2}+\mu^2}{2|\bp||\bp'|}$. In obtaining
(\ref{before_int}) the integration over $p_0'$ has been carried out in the
pole ${\tilde p}_0=\displaystyle\frac{1}{2}\sqsf-E_{p'}$ and, similar to
eq. (\ref{functions}), we define the BS wave function in the continuum as
\begin{eqnarray}
 u_{^1\!S_0}(s_f,|\bp'|)=\frac{g_\1s0({\tilde p}_0, |\bp'|)}{\sqsf-2E_{p'}}.
 \label{u_contin}
\end{eqnarray}
Now, if we restrict ourselves to only one iteration in (\ref{before_int})
taking  the trial function (\ref{u_contin}) as a non-relativistic solution
of the Schr\"odinger equation, e.g. the Paris wave function
$u^{NR}_{^1S_0}(s_f,|\bp'|)$, the BS amplitude is obtained as
\begin{eqnarray}
\phi_\1s0(p_0,|{\bf p}|)=\phi_\1s0^0(p_0,|{\bf p}|) -
\frac{G^{o.i.}(\tilde p_0,|{\bf p}|)}
{\left(\frac{1}{2}\sqsf-E_p\right)^2-p_0^2},
\label{foi}
\end{eqnarray}
where the "one-iteration"  BS vertex $G^{o.i.}(\tilde p_0,|{\bf p}|)$
is defined by
\begin{eqnarray}
G^{o.i.}(\tilde p_0,|{\bf p}|)
&=&\frac{1}{\pi}\,\frac{g_{\pi NN}^2}{4\pi}
\left\{ \left[1-\frac{E_p}{m}\right]
\int\limits_0^\infty dr\, e^{-\mu r}j_0(pr)\,u^{NR}_{^1\!S_0}(r)
\right.
\nonumber\\
&+&
\left .\frac{|\bp|}{m E_p} \int\limits_0^\infty dr\,
\frac{u^{NR}_{^1\!S_0}(r)}{r}\,e^{-\mu r}\,(1-\mu r)\,j_1(pr)
\right\}.
\label{fin_oi}
\end{eqnarray}
From eqs. (\ref{foi}) and (\ref{fin_oi}) 
one can easily find the non-relativistic
analogue of the obtained formulae. The 
free term in eq. (\ref{foi}) together
with the first term in eq. (\ref{fin_oi}) 
reflect the non-relativistic equation
for the $^1S_0$ wave function, while the 
second term in (\ref{fin_oi}) turns out
to be a correction of  purely relativistic origin.

\subsection{Numerical results}
\begin{figure}[hb]  %Fig.7
\epsfxsize 5in
\centerline{\epsfbox{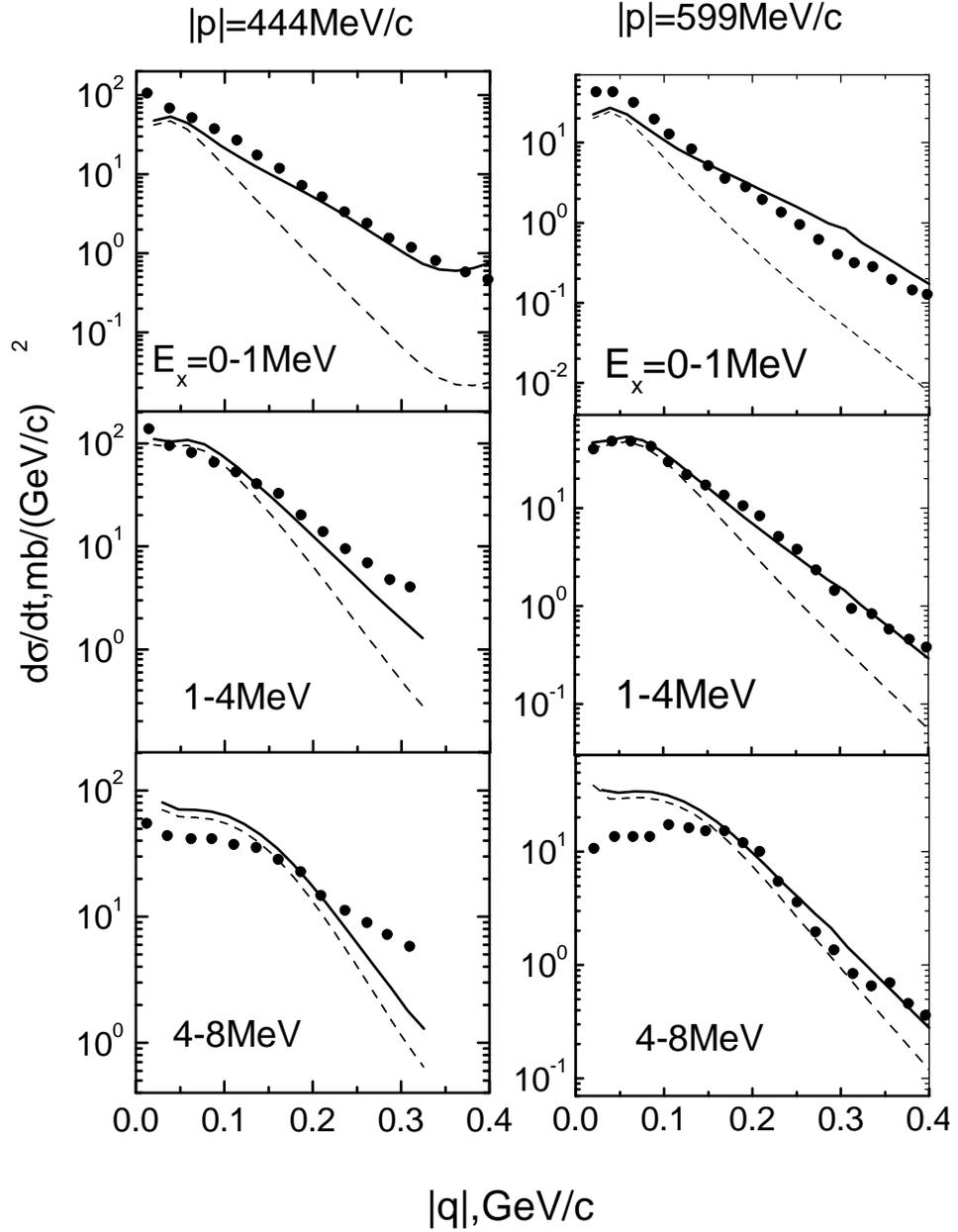}}
\vfill
\caption{ Results of calculations of the differential cross section eq.
 (\ref{sech}) with taking into account the effects of final
 state interaction in $^1S_0$ state (solid curves). Experimental
 data are those from SATURN-II~\cite{kox}, 
the elementary amplitude has been taken
 from  refs. \protect\cite{said,saidpap}. The dashed curves depict
 the results of calculations within the  pure impulse approximation
 (cf. solid curve in Fig. \ref{pict5}).
}
\label{pict7}
\end{figure}

\begin{figure}[hb]  %Fig.8
\epsfxsize 5in
\centerline{\hspace*{-1cm} \epsfbox{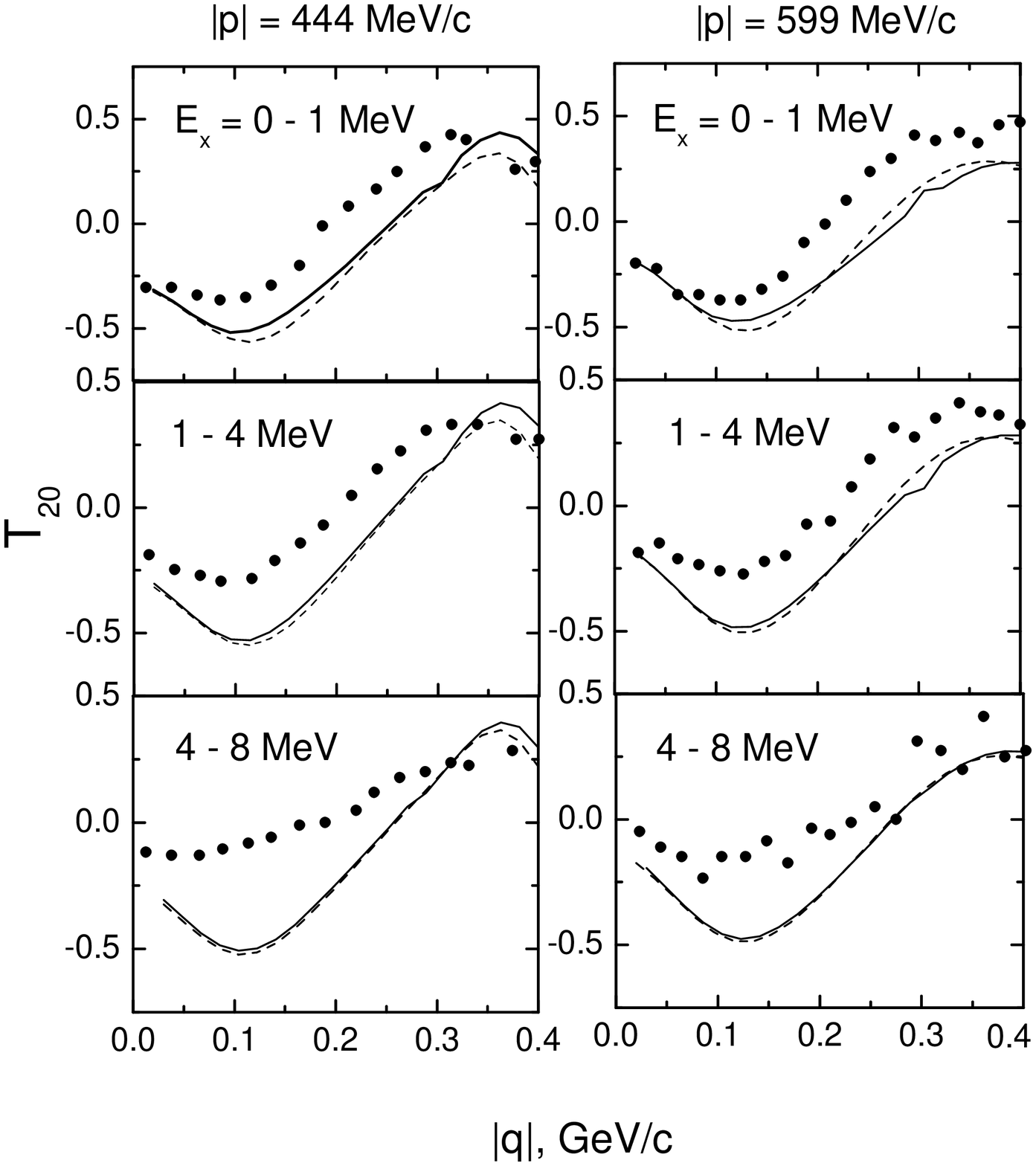}}
\vfill
\caption{The  same as in Fig. \ref{pict7} but for the
 tensor analyzing power (\ref{t20}).
}
\label{pict8}
\end{figure}

In Figs.~\ref{pict7} and \ref{pict8} we 
present results of numerical calculations of
the cross section  and
tensor analyzing power  $T_{20}$ given by eqs. (\ref{sech}),
(\ref{t20}), (\ref{ampF}), (\ref{mel}) and (\ref{foi}-\ref{fin_oi}).
The elementary
charge-exchange amplitude has been taken from ref.~\cite{saidpap} and
the non-relativistic trial function $u^{NR}_{^1S_0}(r)$ as the
solution of the Schr\"odinger equation 
within the Paris potential~\cite{paris_cont}.
The BS $S^{++}$ and $D^{++}$ amplitudes are those from
the  numerical solution \cite{solution}
 obtained with a realistic one-boson exchange interaction.
The dashed  curves in Figs. \ref{pict7} and \ref{pict8} correspond to
results within the relativistic impulse approximation, while the solid lines
depict results with taking into account the final state interaction in
one-iteration approximation.
 It is seen that in all three energy bins 
the agreement with data for the cross section
 is essentially improved. This 
concerns especially the range $1\leq E_x\leq 4$ MeV. For
the  energy bin close to zero there is still 
a disagreement with data at low transferred
 momenta which probably may be related to the fact that in our calculations
 we have not taken into account the Coulomb interaction within 
the $pp$ pair. For
  higher excitation energies ($E_x\sim 8$ MeV), other partial waves (e.g.
 triplet state) in the $pp$ final state contribute and, within the adopted
 assumptions, one may expect only semi-quantitative agreement with data. From
 Fig. \ref{pict7} one may conclude that at low excitation energies
 the supposed mechanism for the reaction (\ref{reaction}) 
(i.e. charge-exchange subprocess with  
interaction in $^1S_0$ state of the $pp$ pair
 in the continuum) seems to be correct. Moreover, from a comparison of
 the left and right panels in Fig.~\ref{pict7} one may expect that the
 higher initial energy the larger kinematical region where the mechanism
 holds. Fig. \ref{pict8} demonstrates that the tensor analyzing power
 is less sensitive to final state interaction effects. As a matter of fact,
 the tensor analyzing power (\ref{t20}),
 being a ratio of non-diagonal products of partial amplitudes
 to the  diagonal ones, serves as a measure of the quality of parameterization
 of partial amplitudes and their mutual relative phases.
 This has been pointed out
 in a series of publications (see e.g. refs. \cite{inclusive,ourprc}), where
 a good simultaneous description of
 cross sections and $T_{20}$ in reactions of the deuteron
 break-up or elastic scattering of protons, is still lacking.
 Nevertheless, since in the process (\ref{reaction}) the
 behavior of the partial amplitudes
(\ref{eq3}), as seen from eq. (\ref{mel}),
 is mostly governed  by the elementary
charge-exchange amplitudes, an experimental investigation
of the
tensor analyzing power $T_{20}$ in reactions
of the type (\ref{reaction}) can essentially supplement  data on
the $NN$ charge-exchange amplitudes at high energies.

\begin{figure}[hb]  %Fig.9
\centerline{\epsfbox{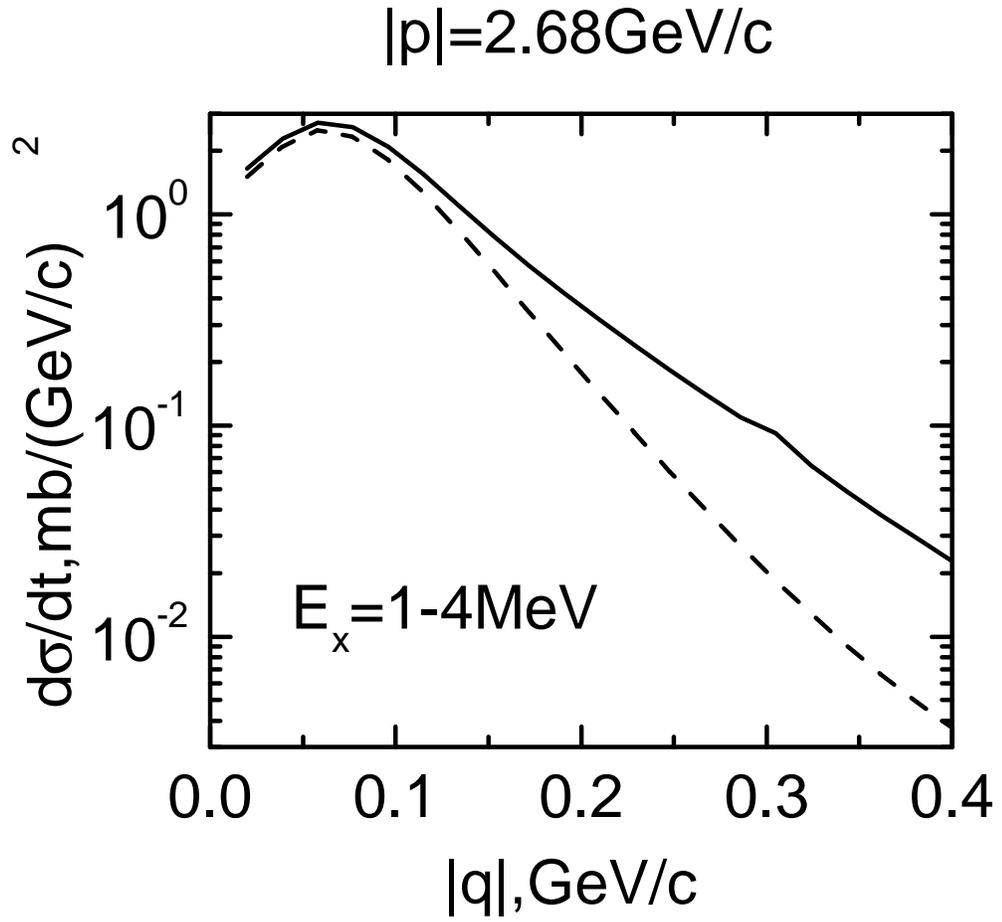}}
\vfill
\caption{ Results of calculations of the differential cross section
 (\ref{sech}) with taking into account the effects of final
 state interaction in $^1S_0$ state (solid curves). Kinematical conditions
 correspond to those proposed in \cite{cosy_proposal} for experiments 
at COSY.
Elementary amplitude 
 from  refs. \protect\cite{said,saidpap}.
 The dashed curves depict  results 
of calculations within the  pure impulse approximation
 (cf. solid curves in Fig.\ref{pict3}).
}
\label{pict9}
\end{figure}

\begin{figure}[hb]  %Fig.10
\centerline{\epsfbox{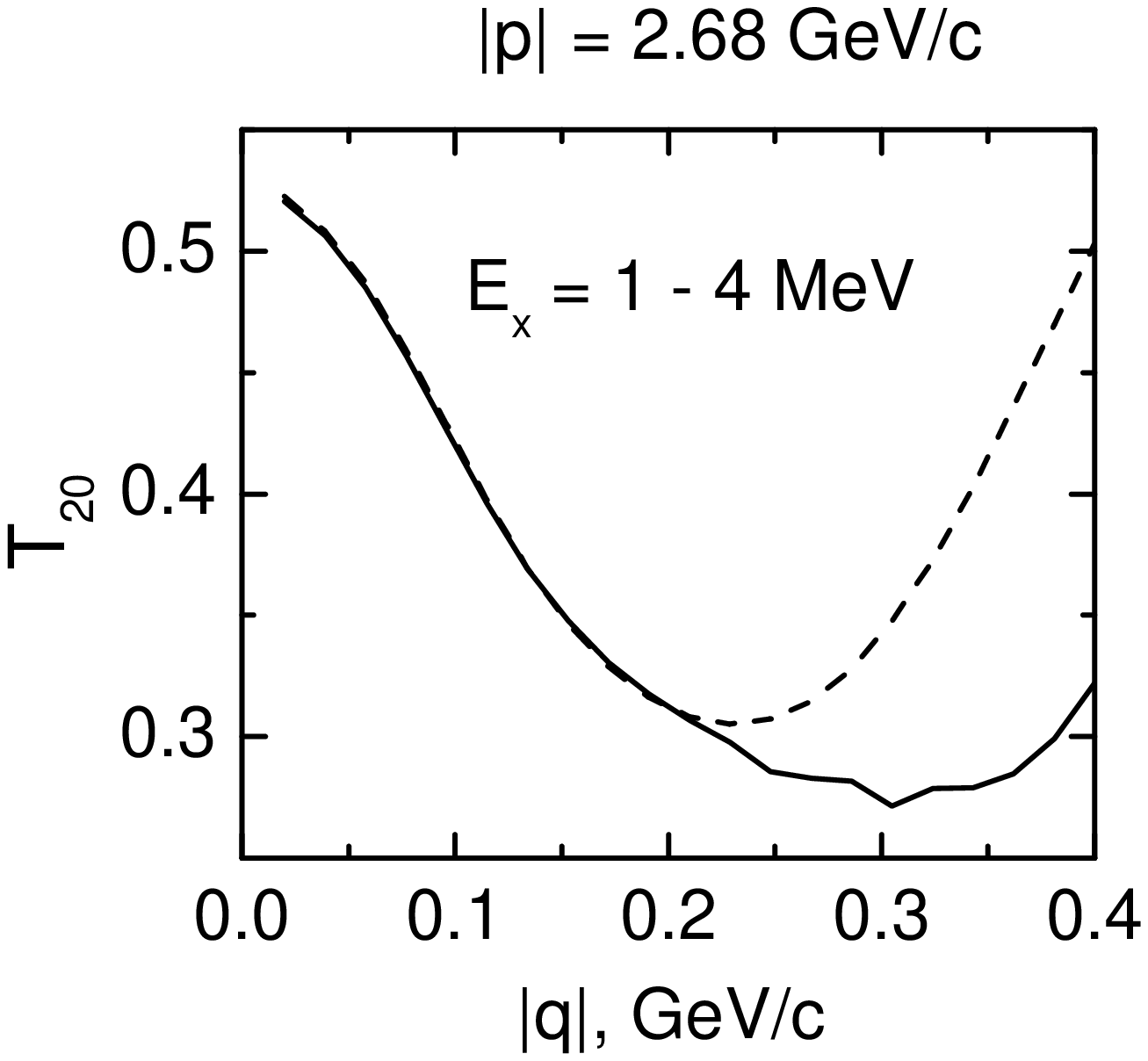}}
\vfill
\caption{The  same as in Fig. \ref{pict9} but for the
 tensor analyzing power (\ref{t20}).
}
 \label{pict10}
\end{figure}

In Figs.~\ref{pict9} and \ref{pict10} we present the predicted cross
section and tensor analyzing power at high energies relevant for 
COSY and Dubna accelerator.
It is immediately seen that the cross section is substantially
decreasing with increasing energy, nevertheless it remains large
enough to be experimentally easily accessible. Another peculiarity of the
studied process at relativistic energies is that the tensor analyzing
power $T_{20}$ does not change the sign remaining positive in a large
kinematical region, in contrast to lower energies (cf. Fig.
\ref{pict8}).
 Note again, that in the above calculations 
the vector polarization of the deuteron
is strictly zero.

From the performed analysis one can conclude that there is a
kinematical region for the excitation energy,  $E_x <5$ MeV, and
transferred momentum, $|{\bf q}| \le\, 0.3 - 0.4$ GeV/c
(i.e. the COSY~\cite{cosy_proposal} kinematics), for which
the mechanism of the reaction (\ref{reaction}) is fairly well described
within the spectator approach by an elementary $pn$  charge-exchange
subprocess, for active nucleons, with detection of the $pp$ pair in
the $^1S_0$ final state.
Our covariant approach agrees with previous non-relativistic
calculations and allows for predictions of the cross sections and
polarization observables at intermediate and relativistic energies,
in particular, for  kinematical conditions which are realized at COSY.
The predicted cross sections of the process and the tensor analyzing power
$T_{20}$ are large enough to be
used, in a large range of initial energies,
for determining properties of the polarized deuteron, provided
experimentally one simultaneously detects a vanishing  vector polarization
of deuterons.

\section{Summary} \label{gl4}

In summary, the performed covariant analysis of the
reaction  $\vec D(p,n)pp$ with the two final protons in a $^1S_0$ state
allows us to conclude that, as in the non-relativistic limit, such a process
can be used as an effective deuteron polarimeter also
at relativistic energies, in particular, at the range covered by 
COSY at J\"ulich and upgraded Dubna accelerator. Additional
information about the elementary charge-exchange amplitude at
high energies can be obtained from precision data with known
deuteron polarization.

\section{Acknowledgments}
We thank R. Arndt and I. Strakovsky for providing information about
the use of their excellent SAID complex of programs.
This work was performed in parts during the visits of S.S.S and L.P.K.
in the Forschungszentrum Rossendorf, Institute of Nuclear and Hadron Physics.
The support by the program "Heisenberg-Landau" of JINR-FRG
collaboration and the grants BMBF 06DR921, WTZ RUS 98/678 and RFBR 00-15-96737
are gratefully acknowledged.
%See \ref{sect:aa}

%\section{Appendix}
\appendix\section{Deuteron state}
\label{sect:aa}

%\setcounter{equation}{0}

%\begin{enumerate}
%\item 
%8.1.
The BS amplitudes of the deuteron, eq. (\ref{deu}),
in the deuteron center of mass explicitly read
\begin{eqnarray} \Psi^M_{S^{++}}(k)  &=&
{\cal N}(\hat k_1+m )\frac{1+\gamma_0}{2}\hat\xi_M(\hat k_2-m) \psi_S
(k_0,|\bk|), \label{psis1}\\[2mm] \Psi^M_{D^{++}}(k) &=&  -\frac{{\cal
N}}{\sqrt{2}} (\hat k_1+m )\frac{1+\gamma_0}{2} \nonumber\\[2mm]
&\times&
\left (
\hat\xi_M +\frac{3}{2|\bk|^2} (\hat k_1-\hat k_2)(k\xi_M)\right )
(\hat k_2-m)
\psi_D (k_0,|\bk|),\label{psis}
%\label{psid},
\end{eqnarray}
where
$k_{1,2}$ are on-mass shell 4-vectors,
\begin{equation}
k_1=(E,\bk),\quad k_2=(E,-\bk),\quad k=(k_0,\bk),\quad
E=\sqrt{\bk^2+m^2},
\label{per}
\end{equation}
and $\psi_{S,D} (k_0,|\bk|)$ are the partial scalar amplitudes, related to
the corresponding partial vertices as
\begin{eqnarray}
\psi_{S,D} (k_0,|\bk|)=
\displaystyle\frac{G_{S,D} (k_0,|\bk|)}
{\left(\displaystyle\frac{M_D}{2}-E\right)^2-k_0^2}.
\nonumber
\end{eqnarray}
In Eqs.~(\ref{psis1}-\ref{psis}) the normalization factor is
${\cal N}=\displaystyle\frac{1}{\sqrt{8\pi}}\displaystyle\frac{1}{2E(E+m)}$.

\section{$^1S_0$ state}
\label{sect:bb}

%8.2.
%\item
 Different representations of the BS amplitude
 $^1S_0$  in the continuum have been studied in details in ref.~\cite{nashi},
 where the reader may found the most general expressions for the covariant
 amplitudes $b_i$, eq. (\ref{covarj0}),
 in terms of  the partial amplitudes 
$\phi_j$ in the center of mass of the $pp$ pair.
 Since in the present paper we consider only the
 $++$ component of the $\phi_i$, we are left with 
 one invariant function $b$, which is taken to be  $b_4$.
Then the explicit expressions for the kinematical coefficients
$C_1, C_2, C_3$ in eq. (\ref{mel}) can be cast in the form
\begin{eqnarray}
\nonumber
&&C_1=(\bp_p\cdot\bk)\,A_1-(E+m)\,(p_p^0+m)\,A_2-(\bq\cdot\bp_p)\,(E+m)\,A_3-
(\bq\cdot\bk)\,(p_p^0+m)\,A_4,\\ \nonumber
&&C_2=\frac{3}{\sqrt{2}(E-m)}
\,[\,(E-m)\,A_1-(p_p^0+m)\,A_2
- ((2\bk-\bq)\cdot\bq)\,A_3],\\ \nonumber
&&C_3=A_1-(E+m)\,A_3+(p_p^0+m)\,A_4,
\end{eqnarray}
where
\begin{eqnarray}
&&A_1={\cal K}\,[P_0'\,(p_p^0+2m)+(\bq\cdot\bp_p)]
-1,\label{d1}\\
&&A_2={\cal K}\,[P_0'\,(p_p^0-2m)+(\bq\cdot\bp_p)]
-1,\label{d2}\\
&&A_3=-{\cal K}\,(2m-p_1^0+p_p^0)\label{d3},\\
&&A_4=-{\cal K}\,(2m+p_1^0-p_p^0),\label{d4}
\end{eqnarray}
with the invariant coefficient ${\cal K}=\displaystyle\frac{1}{e\sqrt{s_f}}$
(see ref.~\cite{nashi}).
The  4-vectors $p_1$ and $p_p$
 are defined by
\begin{eqnarray}
&&\nonumber p_1=\frac{1}{2}P'+k-\frac{1}{2}q=(p_1^0,\bp_p),\\
&&\nonumber p_p=(p_p^0,\bp_p),
\quad p_p^0=\sqrt{m^2+\bp_p^2},\quad \bp_p=\bk-\bq.
\end{eqnarray}
Observe, that $p_p$ represents an  on-mass shell vector.
Note also, that in  the relativistic impulse  approximation,
since $p_1^0=p_p^0$,
 eqs. (\ref{d1}-\ref{d4}) are substantially simplified.

\section{Relativistic spin transformations}
\label{sect:cc}

%Eshe formulae8.3.
%\item
By definition, a state with given momentum $\bp$ and helicity $\lambda$
in a frame of reference $\CO$ is that obtained by
a Lorenz transformation of a state with given spin projection
$s_z$ from the rest system
$\CO_{rest}$ to $\CO$, i.e.:

\begin{equation}
|\bp;\lambda\rangle\equiv | \stackrel{0}{p},s,s_z\rangle_\CO,
\end{equation}
where $\stackrel{0}{p}=(m,0,0,0)$. As usual, a Lorenz transformation $h[\bp]$
is presented by a sequence of two operations:
a boost along the $z$ axis, $l_z(v)$, where $v$ is
the speed of the state in $\CO$, and a rotation from $z$ direction to the
direction of $\bp$, i.e.
$\CO=r^{-1}(\phi,\theta,0)l_z^{-1}(v)\CO_{rest}$.

Let us suppose now
that one has a state $|\bp;\lambda\rangle$ given in the frame $\CO$ and
one wishes to know how it reads in another frame  $\CO'$ obtained
by a Lorenz transformation
$l$ on $\CO$
\begin{equation}
|\bp;\lambda\rangle_{\CO'} = U(l^{-1})|\bp;\lambda\rangle.
\end{equation}
From the definition of the helicity states one has
\begin{equation}
U(l^{-1})|\bp;\lambda\rangle =
U(l^{-1}) U(h[(\bp)])| \stackrel{0}{p};\lambda\rangle ,
\label{b3}
\end{equation}
where $h[\bp]$ is the corresponding 
Lorenz transformation $\stackrel{0}{p}\to p$.
Then multiplying eq.~(\ref{b3}) by unity,
$U(h[(\bp')])U^{-1}[h[(\bp')]=1$, where $h[\bp']$
is the helicity transformation that defines a state
$|\bp';\lambda\rangle =U(h[(\bp'])| \stackrel{0}{p},\lambda\rangle $ with
$\bp'$ being the same vector as obtained by transforming $\bp$ from
$\CO$ to $\CO'$,
one obtains:
\begin{equation}
U(l^{-1})|\bp;\lambda\rangle =
U(h[(\bp']){\cal R}| \stackrel{0}{p},\lambda\rangle ,
\end{equation}
where
${\cal R}=U^{-1}[h[(\bp')]U(l^{-1})U[h(\bp)]$ is the 
sequence of transformations
$ \stackrel{0}{p}\to p \to p' \to  
\stackrel{0}{p}$, i.e. nothing but a rotation.
Then,
\begin{equation}
|\bp,\lambda\rangle_{\CO'} = D^{(s)}_{\lambda\lambda'}(\omega)
|\bp',\lambda'\rangle,
\end{equation}
where $\omega$ is a set of Euler angles describing the rotation.
In case when the Lorenz transformation is a 
simple boost along the $z$ direction
with the speed $\beta$, then $\omega$ is just an angle, describing a rotation
about the $Y$ axis,
\begin{eqnarray}
\cos\omega=\cos\theta'\cos\theta+ \gamma\sin\theta'\sin\theta,
\end{eqnarray}
with $\gamma=1/\sqrt{1-\beta^2}$, and $\theta, 
\theta'$ are the polar angles of $\bp$
in the systems $\CO$ and $\CO'$, respectively.
This is known as Wick helicity rotation,
contrary to Wigner's canonical spin rotation.
In our case, the relevant $z$ axis is the one along the  direction of
 $(\bk+\bp)$.
Then, obtaining the helicity amplitudes in the laboratory frame
we need an additional
rotation to change from the helicity basis to the spin projections.
%\end{enumerate}

%%%%%%%%%%%%%%%%%%% references
%%%%%%%%%%%%%%%%%%%%%%%%%%%%%%%%%%%%%%%%

\newpage
%%                            ---------  FIGURES ----------------------
%%                            --                                      -

\newpage

\begin{figure}[hb]   %Fig2.
\centerline{\epsfbox{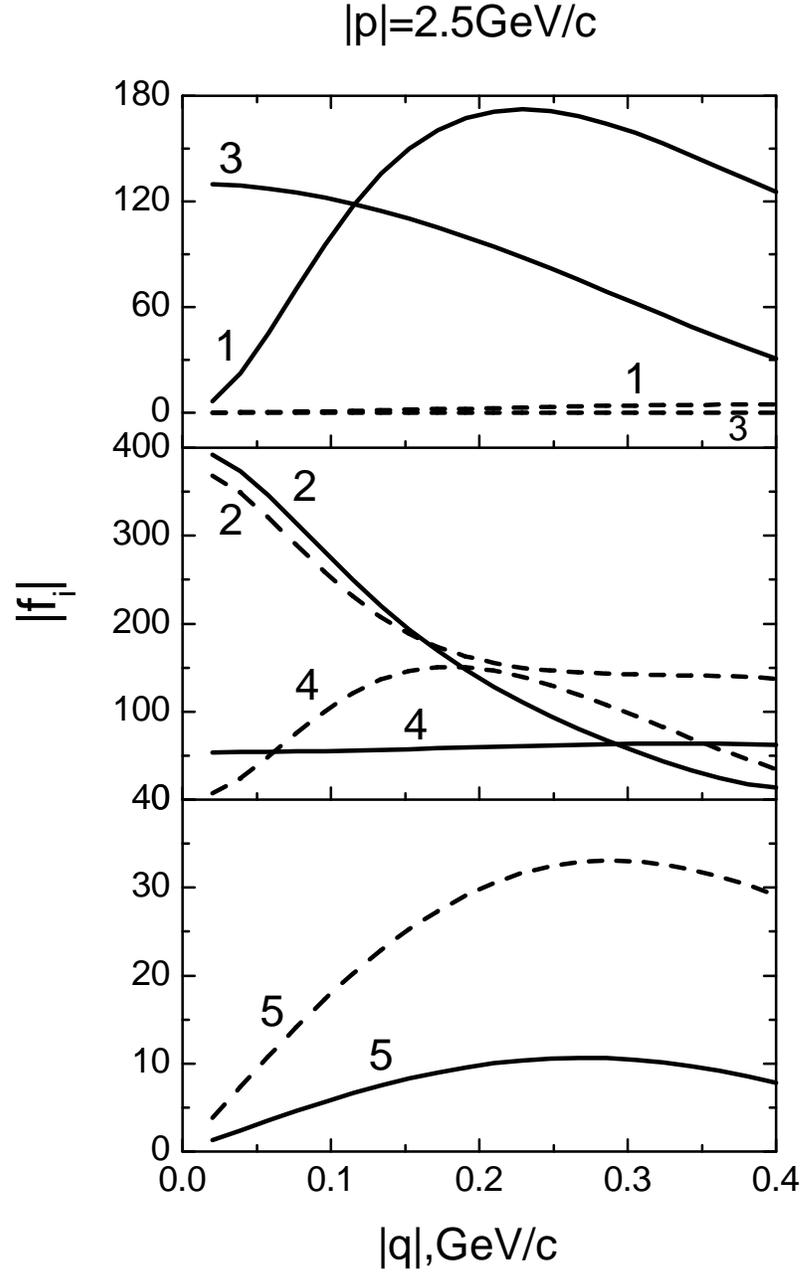}}
\vfill
\caption{Partial helicity amplitudes of eq. (\protect\ref{helicity}) vs. the
transferred momentum $|{\bf q}|$
for different parameterizations.  Solid curves correspond to the
partial wave analysis of \cite{said,saidpap},
while the dashed curves are the high energy parameterization given
in \cite{diu}. The amplitudes are dimensionless.
}
\label{pict2}
\end{figure}

\newpage

\newpage

\newpage

\newpage

\newpage

\newpage

\newpage

\newpage
\end{document}